\documentclass[manuscript]{acmart} 

\AtBeginDocument{%
  }

\acmSubmissionID{5810}

\usepackage{longtable}
\usepackage{enumitem}
\usepackage{xspace}
\usepackage{listings}
\usepackage[most]{tcolorbox}
\usepackage{enumitem}
\usepackage{longtable}
\usepackage{array}
\usepackage{booktabs}

\newcommand{\app}{\textsc{AttriFy}\xspace}

\newcommand{\icon}[1]{\raisebox{-0.2em}{\includegraphics[height=2.5ex]{#1}}}
\newcommand{\prt}[1]{\textit{Part~#1}}

\authorsaddresses{}

\newcommand{\smallparagraph}[1]{\refstepcounter{paragraph}\noindent\textbf{#1\ ---}\label{par:\theparagraph}}

\usepackage{microtype}
\newtcolorbox{promptbox}[1][]{
  colback=gray!4,
  colframe=gray!60,
  boxrule=0.4pt,
  arc=1pt,
  left=6pt, right=6pt, top=4pt, bottom=4pt,
  fonttitle=\bfseries\small,
  title=#1,
  breakable
}

\lstdefinestyle{promptcode}{
  basicstyle=\scriptsize\ttfamily,
  breaklines=true,
  breakatwhitespace=false,
  columns=fullflexible,
  keepspaces=true,
  showstringspaces=false,
  frame=none,
  xleftmargin=0pt,
  aboveskip=0pt,
  belowskip=0pt,
  breakindent=0pt,
  postbreak=\mbox{\textcolor{gray}{$\hookrightarrow$}\space}
}

\begin{document}

\title[Interactive Modeling of Soft Attributes]{\textit{Let Me Clarify}: Interactive Modeling of Soft Attributes}

\author{Steffen Holter}
\email{steffen.holter@inf.ethz.ch}
\affiliation{%
  \institution{ETH Zurich}
  \city{Zurich}
  \country{Switzerland}
}

\author{Rita Sevastjanova}
\email{rita.sevastjanova@inf.ethz.ch}
\affiliation{%
  \institution{ETH Zurich}
  \city{Zurich}
  \country{Switzerland}
}

\author{Michaela Benk}
\email{michaela.benk@uzh.ch}
\affiliation{%
  \institution{University of Zurich}
  \city{Zurich}
  \country{Switzerland}
}

\author{Mennatallah El-Assady}
\email{melassady@ai.ethz.ch}
\affiliation{%
  \institution{ETH Zurich}
  \city{Zurich}
  \country{Switzerland}
}

\renewcommand{\shortauthors}{Holter et al.}

\begin{abstract}

People commonly express preferences through soft attributes such as ``\textit{is interesting},'' which are difficult to operationalize for computational preference modeling. Latent representations can obscure how such attributes are interpreted, limiting transparency and user control. We formalize softness as the diversity of valid interpretations a description permits and provide empirical support for this characterization. Our proposed measure estimates softness to guide clarification and determine when to generate computational approximations. This is instantiated in \app, a mixed-initiative system that translates descriptions into executable Python functions over text corpora. An interactive dashboard connects the dialogue to proposed functions and computed values, enabling users to inspect, accept, reject, and refine interpretations. An exploratory within-subjects study with 32 participants compares our softness-guided clarification approach with an LLM baseline in supporting interpretable attribute construction. This work positions softness as a basis for constructing explicit, user-steerable attribute dimensions for downstream preference elicitation and decision-making.

\end{abstract}

\ccsdesc[500]{Human-centered computing~Interactive systems and tools}
\ccsdesc[500]{Human-centered computing~Natural language interfaces}

\keywords{soft attributes, interactive attribute construction, preference elicitation, human-AI collaboration, chat-based interfaces}

\begin{teaserfigure}
  \includegraphics[width=\textwidth]{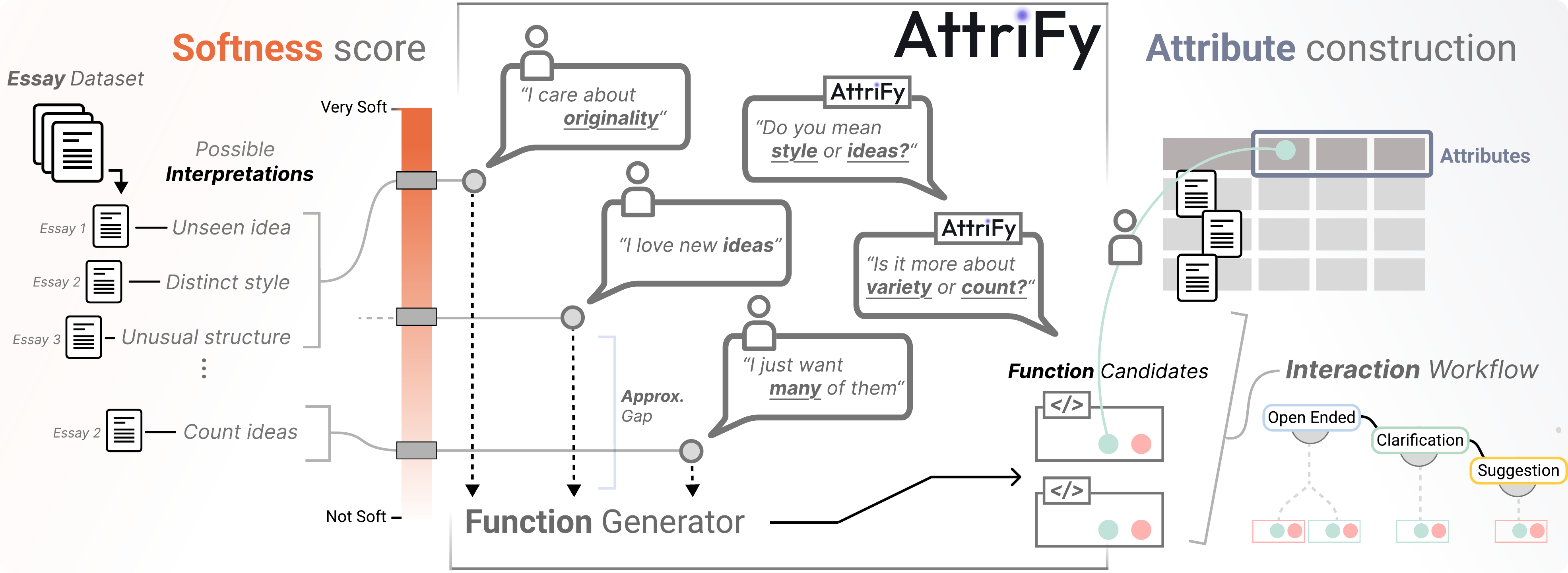}
  \caption{\textbf{Overview of AttriFy:} Our system helps users model soft attributes through clarification, turning broad concepts such as ``originality'' into executable attributes that reflect their intended meaning. The figure illustrates our two main contributions: a softness measure that captures the extent to which a description admits multiple interpretations (left) and an attribute construction approach that guides users through a three-step workflow of open-ended questions, clarification, and user suggestions (right).}
  \Description{An overview of AttriFy’s workflow, shown in three parts. On the left, the ``Softness score'' illustrates how a user attribute can have multiple possible interpretations. Using an essay dataset as an example, ``originality'' might refer to different interpretations. A vertical scale ranges from ``Very soft'' at the top to ``Not soft'' at the bottom, with broader and more ambiguous interpretations appearing higher on the scale and more concrete interpretations. In the middle, AttriFy gradually clarifies what the user means through conversation. The user first says, ``I care about originality,'' then narrows this to new ideas and finally to wanting many ideas. As the meaning becomes more specific, the corresponding softness decreases. Dashed arrows connect each stage to a ``Function Generator,'' with the distance representing the approximate gap between the user’s description and a measurable function. On the right, the generator proposes candidate attribute functions that the user can accept or reject. Accepted functions become columns in an attribute table containing values for each essay. The workflow at the bottom summarizes the three interaction stages: Open Ended, Clarification, and Suggestion.}
  \label{fig:teaser}
\end{teaserfigure}

\maketitle

\section{Introduction}
\label{sec:intro}

Natural language is increasingly used to elicit user preferences in automated decision-making systems~\cite{kostric_2021_soliciting,li_2025_gate,andukuri_2024_stargate}. 
Existing methods capture natural language preferences either implicitly within a latent space~\cite{li_2025_gate} or explicitly by mapping input to a fixed set of static attributes~\cite{kostric_2021_soliciting}. 
However, natural language descriptions are inherently \textbf{soft}: they exhibit a high degree of semantic imprecision and admit multiple defensible interpretations without further specification~\cite{balog_2021_interpretation}. 
In a latent-space paradigm, a single soft descriptor may conflate distinct semantic dimensions and make downstream rankings uninterpretable. 
For example, evaluating essays based on ``originality'' can obscure whether a score reflects conceptual novelty, stylistic deviation, or structural divergence. 
Mapping language directly to explicit attributes increases operational transparency, yet no direct mapping exists between soft language and formal criteria. 

This lack of a direct mapping stems from the fact that a soft description leaves open multiple valid ways to construct a computable function.
Resolving such openness is not simply a matter of recovering a fixed meaning presumed to exist in the user's mind, nor is it addressed by typical forms of linguistic ambiguity, such as lexical polysemy, semantic mapping, or vagueness thresholding~\cite{yadav_2021_comprehensive}.
These approaches resolve existing structures, for instance, polysemy selects among word senses, semantic mapping links expressions to fixed ontologies, and vagueness sets boundaries on scalar predicates. 
Soft descriptions, however, pose a different challenge: their interpretation is inherently subjective and user-dependent, and therefore cannot be fully determined from linguistic context alone. 
Moreover, users may be unable to articulate what they mean upfront, discover relevant distinctions only after seeing how their description is applied, or revise their preferences during the process.
Constructing an attribute from a soft description is therefore an interactive specification problem: the user and the system must reach an operational meaning that can be inspected and applied consistently.\looseness=-1

One way to reduce the softness of a user's description is by actively clarifying their intent through targeted questions~\cite{aliannejadi_2019_asking} – an approach explored in prior conversational preference elicitation frameworks~\cite{andukuri_2024_stargate}.
Clarification, however, incurs interaction overhead, making it unfeasible to query every possible interpretation. While some descriptions provide a sufficiently clear basis for computation, others leave open ambiguities that materially alter downstream outcomes. 
Effective systems must therefore identify when a description is genuinely soft so that clarification is requested only when its value outweighs the cost—an objective addressed by recent efforts in cost-aware clarification under ambiguity~\cite{zhang_2025_clarify}.

In this paper, we introduce \app, a mixed-initiative visual analytics system for constructing computable attribute functions from natural-language descriptions.
We develop a computational estimate of softness and treat it as a control signal for interactive attribute construction.
Specifically, \app uses estimated softness to identify when a description may benefit from clarification and guides the user through an iterative dialogue to construct an explicit operationalization.
A user-facing agent proposes dataset-grounded refinements that help the user distinguish between competing interpretations, while a modeling agent translates the resulting specification into executable functions over the data.
A visual interface exposes the provenance of this process, allowing users to trace the path from their initial description to the resulting functions and inspect or revise intermediate decisions.
The resulting functions represent explicit operationalizations of the description rather than claims to its uniquely correct or ground-truth meaning.


The paper is structured in the following way. 
After reviewing prior work, we establish the design goals of our interface. Next, to demonstrate that softness can serve as an effective signal, we formalize and validate it through a formative study. 
Our results show that people share an ordering of descriptions by softness, that this ordering tracks the diversity of their interpretations, and that their judgments respond systematically to clarification.
These findings provide an empirical reference for evaluating our computational estimate.
We then present our agent-based workflow and the interface design, concluding with an evaluation of the end-to-end approach comparing \app to a baseline.
In this evaluation, we assess whether the resulting attributes reflect users' judgments and discriminate meaningfully among items.




To summarize, this work makes three contributions:
\begin{enumerate} [leftmargin=12pt,labelsep=0.3em]
\item \textbf{Softness:} We define and empirically ground softness as interpretive diversity and conduct a formative study showing that perceived softness corresponds to the number of human interpretations and changes systematically through clarification.
\item \textbf{\app:} We introduce a mixed-initiative system that uses our proposed computational estimator of softness to decide when to clarify and when to generate executable attribute functions. Additionally, dataset-grounded questions and a visual interaction history support users in constructing and refining attributes.
\item \textbf{Evaluation:} We compare \app to an unstructured LLM baseline in a within-subjects study, examining perceived support for clarifying intent, interaction experience, and interaction patterns.The results characterize where softness-guided clarification changes the construction process and where it does not.
\end{enumerate}

\section{Related Work}

Below, we review related work along four core axes central to our approach: capturing user preferences, managing ambiguity in natural language, iteratively clarifying user intent, and designing visual interactive chat-based interfaces.


\subsection{Capturing User Preferences}
Several lines of work have attempted to operationalize subjective descriptors for use in computational pipelines. 
\citet{li_2025_gate} investigate how language models can actively elicit and infer human preferences through interactive questioning.
Concept activation vectors are used to model soft attributes as directions in a learned embedding space~\citep{balog_2021_interpretation,gopfert_2024_discovering}. The approach has been extended to active critiquing - given a set of items, the technique determines the soft attributes about which to elicit user feedback~\citep{biyik_2023_preference}.  
More interactive approaches employ multi-turn Bayesian elicitation to iteratively refine user preferences~\citep{austin_2024_elicitation}, address noisy or inconsistent user feedback~\citep{escamocher_2025_interactive}, as well as use function approximation to obviate the need for specific modeling assumptions~\citep{martin_2024_model}. 
Also item pairwise comparisons are used to learn preference models, e.g., in the latent space of an autoencoder~\citep{eichelbeck_2026_supporting}.
Instead of yielding transparent, directly computable attribute definitions, these methods produce opaque representations that users can neither easily inspect nor adjust.
Other systems focus on developing preference elicitation methods that can accommodate soft attributes by combining item-based and attribute-based feedback~\citep{biyik_2023_preference}. 
Across these approaches, a common limitation persists as they provide limited insight into how a subjective description is translated into measurable criteria, which constrains trust, user control, and suitability for practical decision support.
Interactive tools for constructing LLM evaluation criteria~\cite{shankar_2024_validators} come closest, showing that users revise their criteria as they inspect outputs, but they assume a criterion can be stated and then aligned, whereas we address the prior question of whether a description is specified enough to be operationalized.


\subsection{Ambiguity in Natural Language}
Ambiguity is a hallmark of natural language. 
Speakers routinely omit details that are inferable from context, which enables concise communication but complicates computational interpretation. 
In machine translation, for example, a single source sentence can admit multiple valid renderings in the target language~\citep{fernandes_2023_translation}. 
Even among human annotators, seemingly well-defined examples can provoke disagreement~\citep{nie_2020_learn}, a phenomenon frequently attributed to underlying linguistic ambiguity~\citep{liu_2023_afraid}. These observations highlight the importance of establishing \emph{common ground} (i.e., a shared understanding between interlocutors) before a task can be solved efficiently~\citep{shaikh_2024_grounding}. The same applies when a human interacts with a chatbot.

Research on natural-language ambiguity has traditionally focused on word-sense and syntactic disambiguation, while recent work increasingly addresses ambiguity in open-ended questions and dialogue, especially in chat-based interfaces. 
It has been shown that LLMs often struggle to explicitly represent ambiguity~\citep{liu_2023_afraid}, although the spread of their sampled outputs has been used as a proxy for it~\cite{kuhn_2023_semantic,cole_2023_selectively}.
A recent survey~\citep{tanjim_2025_disambiguation} shows that work has shifted toward interactive disambiguation approaches, where systems detect ambiguity and aim to resolve it. 
These include approaches like a pipeline that aligns LLMs to manage ambiguous queries by leveraging their own assessment of ambiguity (i.e., perceived ambiguity)~\citep{kim_2024_aligning}, or a `round-trip' approach, where the model generates possible question–answer interpretations, then uses them to generate additional interpretations that may have been missed, and verifies the resulting question–answer pairs to produce the final disambiguated output. 
At the same time, more benchmarks~\citep{sarwar_2026_clarity,min_2020_ambigqa} are created to improve and evaluate the disambiguation methods developed.

According to Poesio, ``ambiguity is one of the most important aspects of natural language, but one of the most difficult to define and distinguish from related (and unrelated) phenomena''~\citep{poesio_2020_ambiguity}. 
Linguistic work distinguishes ambiguity from vagueness~\citep{kennedy_2007_vagueness}, semantic underspecification~\citep{frisson_2009_semantic}, and subjectivity~\citep{frenda_2025_perspectivist}. Although each of these concepts has a distinct definition, as noted by Poesio, the boundaries between them are not always clear. 
Our work builds on these distinctions to isolate an operational property of natural language input. Specifically, we formalize softness as the diversity of valid interpretations a description permits, directly addressing the boundary between vagueness and underspecification, and provide empirical support for this characterization. We then use softness to trigger targeted clarifications and signal when to generate computational approximations.




\subsection{Iterative Clarification}
Clarification questions have long been considered a remedy for natural language ambiguity.
In open-domain information-seeking conversations, it has been shown that asking a single well-chosen clarifying question improves retrieval performance~\cite{aliannejadi_2019_asking}.
Recent work has investigated how to train or prompt large language models to ask these questions proactively~\citep{andukuri_2024_stargate, zhang_2025_modeling}. 
Yet clarifying behavior remains scarce in practice.
The primary application domain has been information-seeking dialogue, where the system must detect ambiguity in a user query and ask targeted follow-up questions~\citep{wu_2019_proactive, guo_2021_abgcoqa}, with dedicated benchmarks developed to evaluate this capability~\citep{deng_2022_pacific}. 
Related research has introduced proactive chain-of-thought prompting approaches that equip dialogue systems with the ability to plan and take initiative toward a conversational goal~\citep{deng_2023_prompting}.
Several machine learning approaches have been used to teach models to ask clarifying questions.
For instance,~\citet{zhang_2025_modeling} simulate multi-turn interactions with synthetic users, scores candidate clarifying questions based on the accuracy of the final downstream answer, and use Direct Preference Optimization (DPO) to train a model to ask targeted questions only when necessary. \citet{chen_2025_learning} introduce Action-Based Contrastive Self-Training (ACT), a data-efficient method for teaching LLMs conversational policies, particularly the ability to recognize ambiguity and ask appropriate clarification questions. They show that ACT can outperform DPO across several conversational tasks.
Finally, ~\citet{montazer_2025_asking} introduce a training method inspired by diffusion models to teach LLMs how to ask clear, step-by-step questions to uncover user preferences.

In our setting, clarification serves a different purpose: it is not about resolving a factual query which is often the target in prior work but about surfacing latent preferences that shape the meaning of a soft attribute~\citep{li_2025_gate}. 
This makes the problem harder in two respects. 
First, there is no ground truth against which a clarification can be validated, since the attribute is inherently subjective. 
Second, the need for clarification is itself a personalized question (i.e., what is ambiguous to one user may be perfectly clear to another) requiring a mechanism that judges when further interaction is necessary and when it would be redundant.

\subsection{Visual Interactive Chat-based Interfaces}

Users frequently struggle to express multifaceted intentions through flat, linear text feeds~\cite{park_2026_bridging}. 
Standard chat-based interfaces often fail to capture branching information topologies, leading to context loss and inefficient interaction. 
To address these limitations, recent work leverages structured, non-linear representations to externalize context and streamline human decision-making. 
\citet{setlur_2022_you} demonstrate how combining natural language interaction with visual semantics allows users to convey intent and navigate analytical conversations.
They show that a system's conversational interface heavily shapes user expectations regarding its underlying capabilities. 
When responses do not align with these expectations, successful interaction is not simply based on presenting an answer, but rather on providing actionable mechanisms for clarification, refinement, and repair.
Aligning with these insights, systems like AmbigChat~\citep{ma_2025_ambigchat} map vague queries into hierarchical disambiguation trees to refine user intent, allowing users to progressively resolve ambiguities and select among alternative interpretations before receiving an answer. 
Other visual tools such as Needle~\cite{liu_2026_needling} provide multi-level thread visualizations to alleviate cognitive load during complex content triage. 
Similarly, visual trees and graph structures enable users to construct coherent linearized dialogues~\cite{bottona_2026_llmberjack} and manage historical context across multi-turn exchanges~\cite{liang_2024_c5}.
We present an interface that helps users model soft attributes through clarification while maintaining a complete interaction history.

\section{Design Goals}

In this section, we characterize the exact scope of the work, the challenges, and the main requirements that inform our proposed interface and workflow. 

\subsection{Problem Scope} 

Due to the nuanced meaning of several terms, it is important to clarify the scope of this work.
In many practical applications of multi-criteria decision support, such as ranking or recommendation, the central challenge is structuring items according to user preferences.
In this context, we distinguish between \textbf{attributes} and \textbf{preferences}.
An \textit{attribute} is a property that describes each item in a dataset and for which every item has a corresponding value, for example, the flight duration of a travel itinerary.
A \textit{preference}, in contrast, is a user-imposed condition over one or more attributes, such as preferring shorter flight times.
Eliciting or learning these preferences lies outside the scope of this work.\looseness=-1

This work focuses primarily on textual corpora, where the attributes of interest are not explicitly represented and must be constructed instead.
In particular, we focus on \textbf{soft attributes}: subjective, user-dependent properties not explicitly encoded in the data.
Examples include notions such as the novelty of a scientific contribution or the clarity of a piece of writing, which lack a single predefined representation.
Our goal is to operationalize these soft attributes in a measurable form by computing corresponding attribute values for each text item.

\begin{figure}[h]
  \centering
  \includegraphics[width=0.35\linewidth]{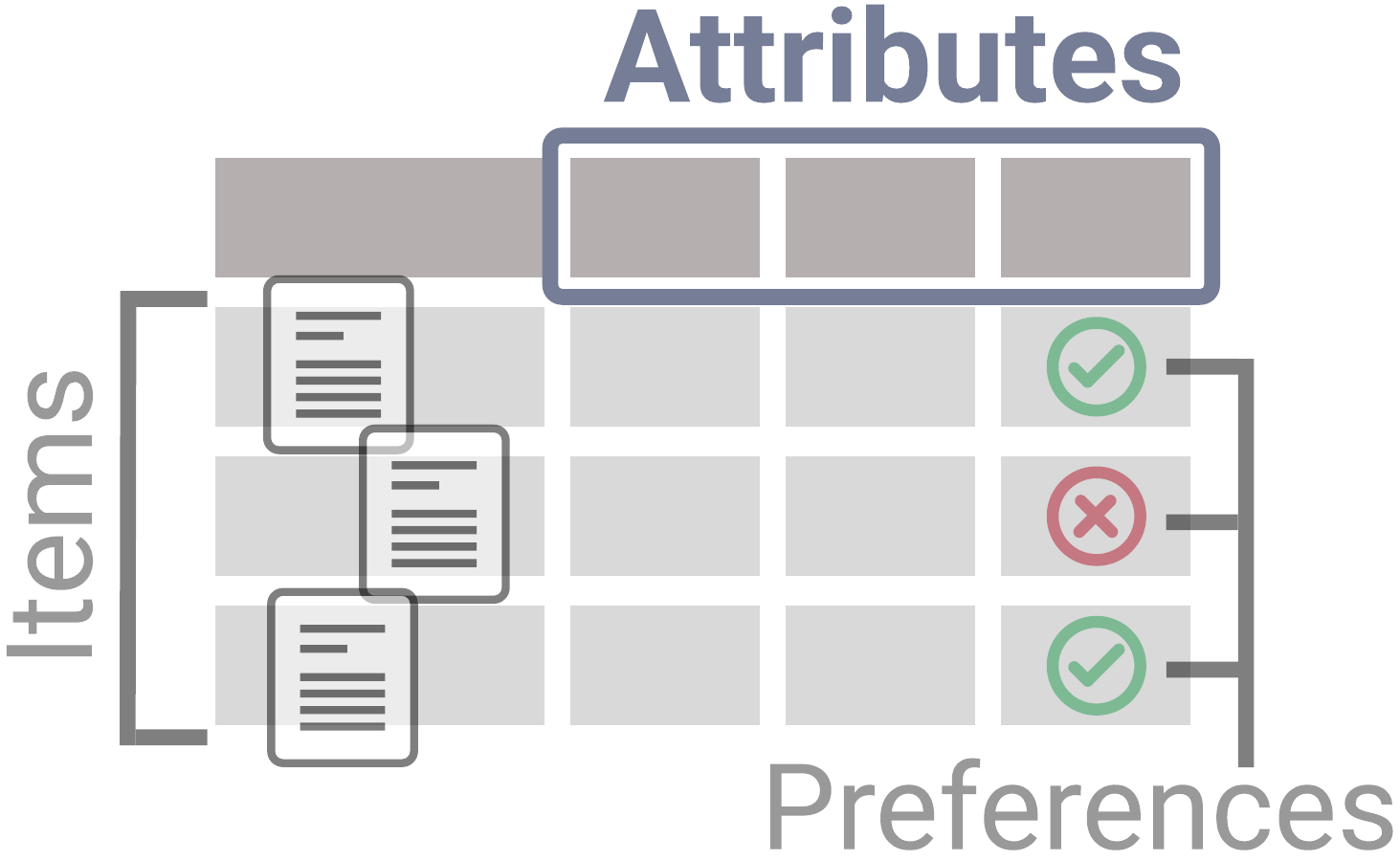}
  \caption{Distinguishing between \textbf{attributes}, preferences and items in a tabular data setting.}
  \label{fig:scope}
  \Description{A simplified view of how items, attributes, and user preferences relate to one another. Each row represents an item, while the columns contain measurable attributes and their values for each item. The final column shows whether each item satisfies the user’s preference based on those attribute values.}
  \vspace{-1em}
\end{figure}

\subsection{User Needs \& Challenges}





Articulating soft attributes and refining how they are operationalized presents several challenges for users.

The construction of attributes is inherently subjective and depends on users' intentions and prior experiences~\cite{golubovikj_2025_rethinking, gopfert_2024_discovering}.
For instance, two users may both express interest in tennis articles for different reasons, with one following the professional circuit and another seeking to improve their own technique~\cite{li_2025_gate}.
The attributes they need to capture these interests will therefore differ.
This highlights why relevant attributes cannot be assumed a priori, but must be grounded in users' intended meanings.
Consequently, predefined or fully automated approaches fail to capture these meanings~\cite{balog_2021_interpretation}, motivating \textit{iterative and interactive refinement of user-described attributes}.

At the same time, users often struggle to explicitly articulate which properties matter to them~\cite{li_2025_gate, zamfirescu_2023_johnny}.
Their initial understanding of relevant attributes may be incomplete, creating a need for support for ideation and exploration of potential attributes.
However, excessive system guidance can introduce bias by steering users toward particular dimensions~\cite{gupta_2024_first}.
This creates a tension between guidance and autonomy, making it necessary to \textit{balance system guidance with user-driven exploration}.

Attribute construction further relies on limited user interactions, providing only partial coverage of the total space of possible attributes.
The selective nature of these interactions can introduce selection bias, as early or salient dimensions are favored while others remain unexplored~\cite{jakesch_2023_cowrite, wadinambiarachchi_2024_effects}.
This concentration on a narrow set of dimensions can produce redundant or highly correlated attributes, limiting the resulting representation's ability to capture diverse user-relevant distinctions~\cite{zhong_2024_explaining,lam_2024_lloom}.
This highlights the need for \textit{diverse and discriminative attribute representations}.

In addition, a user's description of a soft attribute does not necessarily specify how to determine its value for individual items~\cite{balog_2021_interpretation, gopfert_2024_discovering}.
Users' descriptions may leave aspects of their intended meaning implicit, making it difficult to determine how to assess these attributes.
The softer the description, the more assumptions a system may need to introduce to approximate the user's intended meaning with a computational function.
Reducing softness through refinement helps constrain these assumptions.
Therefore, systems must support the \textit{translation of soft attributes into measurable quantities} while preserving users' intended meanings.\looseness=-1

This operationalization, in turn, introduces transparency challenges.
Many existing approaches, such as activation-based methods, lack interpretability~\cite{gopfert_2024_discovering}, making it unclear how user inputs influence the resulting attributes~\cite{radlinski_2022_transparent,balog_2019_transparent}.
This opacity limits users' ability to understand, evaluate, and refine the constructed representations.
As such, systems must provide a \textit{transparent mapping from user input to computed attributes}.\looseness=-1

Finally, transparency must be accompanied by opportunities to act on what users observe.
As users examine attribute values for concrete items, they may identify mismatches with their intended meanings or reconsider their initial definitions~\cite{radlinski_2022_transparent}.
This requires systems to allow users to iteratively adjust, correct, and refine attribute definitions rather than relying on fixed or hard-coded assumptions.
This underscores the importance of \textit{maintaining user control over the attribute construction process}.

\subsection{Description of Requirements}


Based on the challenges outlined above, we formulate the following requirements for a system that supports soft attribute modeling.

\begin{itemize}[leftmargin=15pt,labelsep=0.4em]
\item[R1:] \textbf{Iterative Refinement -} Support the incremental refinement of soft attributes through repeated cycles of inspection and revision.
\item[R2:] \textbf{Balanced Guidance -} Provide guidance that encourages user exploration while limiting system-induced bias.\looseness=-1
\item[R3:] \textbf{Diverse and Discriminative Attributes -} Support the construction of complementary attributes that capture distinct, user-relevant aspects of the data and meaningfully distinguish between items.
\item[R4:] \textbf{Measurability -} Enable the transformation of soft attributes into computable values that reflect users' intended meanings.
\item[R5:] \textbf{Transparency -} Provide interpretable mappings from user input to computed attribute values.
\item[R6:] \textbf{User Control -} Allow users to adapt, correct, and steer the attribute construction process.
\end{itemize}

Together, these requirements motivate an iterative process in which users can inspect and refine how their intended meanings are translated into attribute values. Supporting this process requires an understanding of what makes an attribute soft and how that softness affects its operationalization. The next section develops this conceptual foundation.

\section{Studying Softness}
\label{sec}



Our definition characterizes softness as \textbf{the extent to which a description admits multiple valid interpretations without further specification.}
To ground this definition empirically and inform its computational modeling, we conducted a formative study to examine how people interpret descriptions, judge their softness, and revise those judgments when provided clarification. 
We investigated three research questions:
(1) Is perceived softness associated with the diversity of interpretations a description elicits?
(2) To what extent do people agree in their judgments of softness?
(3) How do these judgments change when a description is clarified or supplemented with additional information? \looseness=-1

Our study focuses on descriptions of middle-school essay attributes, maintaining the application domain introduced earlier.
This setting is familiar and the attributes can be described at varying levels of specificity without requiring specialist knowledge. 
The study comprised three tasks over a common pool of descriptions. 
Participants were first asked to provide interpretations without being introduced to softness, then judged softness through ratings and comparisons, and finally rated descriptions across successive turns of clarifying dialogues. 
Together, these observations provide a human reference for assessing computational estimates of softness.

\begin{figure}[h]
  \centering
  \includegraphics[width=1\linewidth]{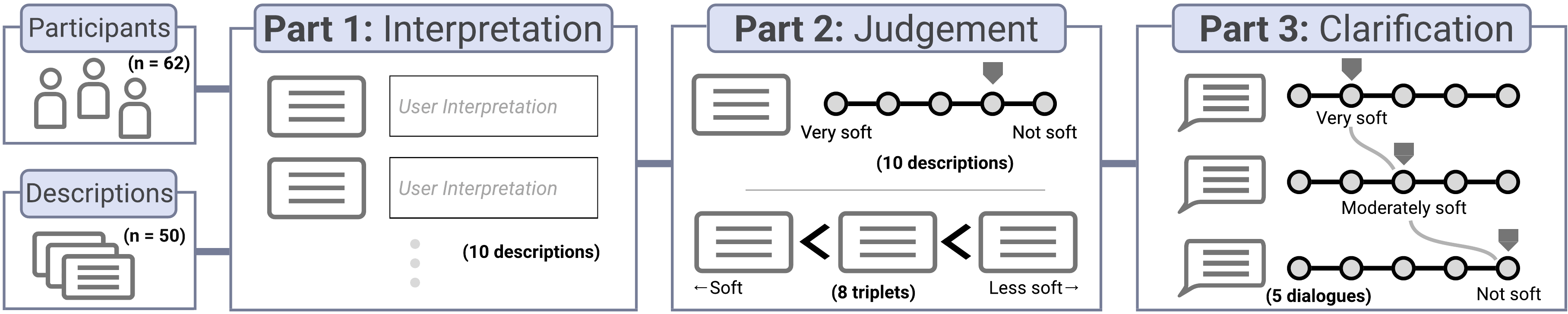}
  \caption{\textbf{Overview of the Study:} (1) Participants interpreted 10 descriptions about essays, (2) rated the softness of 10 descriptions and ranked eight triplets by softness, and (3) rated softness across successive turns of five clarifying dialogues}
  \label{fig:study}
  \Description{An overview of the formative study procedure. The study included 62 participants and a pool of 50 descriptions. In Part 1, Interpretation, each participant saw 10 descriptions and wrote their own interpretation of each one. In Part 2, Judgement, participants rated 10 descriptions on a five-point scale ranging from ``Very soft'' to ``Not soft.'' They also ranked eight sets of three descriptions from softer to less soft. In Part 3, Clarification, participants evaluated five multi-turn dialogues. As each dialogue became more specific, participants rated the softness of the description at different stages, moving from potentially ``Very soft'' toward ``Not soft.''}
  \vspace{0pt}
\end{figure}

\subsection{Study Design}

\smallparagraph{Stimuli}
We constructed a set of descriptions of essay attributes spanning different levels of specificity across dimensions commonly used to assess argumentative writing~\citep{ke_2019_automated}.
For example, ``an essay should include an introduction, a body, and a conclusion'' specifies its organization, whereas ``the ideas should be well argued'' describes the persuasiveness of the content. 
Within each dimension, descriptions ranged from explicit criteria to qualities that different readers could interpret in different ways.
This deliberate variation in specificity guided stimulus construction, while participants' responses supplied the empirical measurements used to model softness.

For the dialogue task, we constructed five types of exchanges with distinct predicted patterns of change in perceived softness.
(1) \textit{Converging} dialogues progressively clarified a description.
(2) \textit{Floor} dialogues began with an already specific description and were expected to remain low in softness, even as more information was provided.
(3) \textit{Plateau} dialogues stopped adding relevant context, either after an initial clarification or from the outset.
(4) \textit{Re-opening} dialogues introduced a new facet mid-way, while (5) \textit{Reversal} dialogues directly contradicted an earlier more concrete statement.
By design the latter two were expected to increase softness.
Including these cases allowed us to examine whether judgments responded to the content of an exchange rather than simply decreasing with additional turns.

We generated candidate descriptions and dialogues using Claude Opus 4.8~\citep{anthropic_2026_opus48}, following the criteria above, and manually removed duplicates and items that did not resemble plausible user descriptions. 
The final pool contained 50 descriptions and 20 five-turn dialogues. 
Generation prompts and full stimuli are provided in Appendix~\ref{app:stimuli}.

\smallparagraph{Participants}
We recruited 70 participants through the crowdsourcing platform Prolific\footnote{https://prolific.com/}.
Participants were required to be fluent in English and have a university-equivalent education, which we used as a proxy for some familiarity with argumentative writing.
After applying our exclusion criteria (Appendix~\ref{app:study-details}), we retained data from 62 participants.\looseness=-1

\smallparagraph{Procedure} 
All participants completed a single web-based session comprising three tasks in a fixed order: (1) interpretation, (2) judgment, and (3) clarification.
Sessions lasted an average of 27 minutes (SD = 8 minutes).
In the \textit{interpretation task}, each participant saw ten descriptions and wrote one sentence per description explaining how they would interpret or measure the described attribute.
At the start of the \textit{judgment task}, we introduced our definition of softness and a five-point rating scale (Appendix~\ref{app:softness-scale}).
Participants then rated ten descriptions individually using this scale and ranked eight sets of three descriptions from least to most soft via a drag-and-drop interface.
In the third \textit{clarification task}, participants viewed five dialogues, each revealed one turn at a time. 
After each turn, they used the same scale to rate the softness of the description based on the information and context provided so far.
No participant both interpreted and rated the same attribute description from the pool.
Item order was randomized within each task.

\subsection{Empirical Analysis of Softness}
For the \textit{interpretation task}, two coders independently grouped participants’ responses to each description.
Responses were assigned to the same group when their interpretations would identify the same set of essays as satisfying the description (Appendix~\ref{app:softness-coding}).
For example, one participant interpreted \textit{``each claim has some evidence behind it''} as ``statements or arguments are followed by facts , statistics, quotes, or specific real-world examples'' while another responded more generally \textit{``look for examples or facts that support each claim.``}
Although phrased differently, these responses were clustered together because they were judged to imply equivalent criteria for evaluating essays.
The coders identified 28 of the 620 responses as unrelated to their corresponding descriptions and excluded them from the analysis.
The two coders’ independent groupings were retained without consolidation.
To account for unequal sample sizes, we used rarefaction to estimate the expected number of distinct groups in a sample of nine responses per description which is the minimum after exclusions.
We then averaged the two coders’ estimates to obtain a single value for each description.
Agreement between the coders’ response groupings was assessed using the Adjusted Rand Index (ARI), and reliability of the averaged rarefied counts was assessed using an absolute-agreement intraclass correlation coefficient (ICC).

For the \textit{judgment task}, we calculated the mean rating for each description and assessed inter-rater agreement across individual ratings using Krippendorff's ordinal $\alpha$~\citep{hayes_2007_coding}. 
Additionally, each ranking of three descriptions was converted into three pairwise comparisons, one for each possible pair. 
For each description, we calculated the proportion of comparisons involving it in which it was ranked softer than the other description.
Spearman’s rank correlations $\rho$ were then used to assess the associations between pairwise comparison scores and mean softness ratings, and between mean softness ratings and rarefied interpretation counts.

For the \textit{clarification task}, we examined rating trajectories across turns and changes from the first to the final turn. 
For dialogues predicted to plateau or reverse direction, we also considered when changes occurred within the exchange. 

\begin{figure}[h]
  \centering
  \includegraphics[width=1\linewidth]{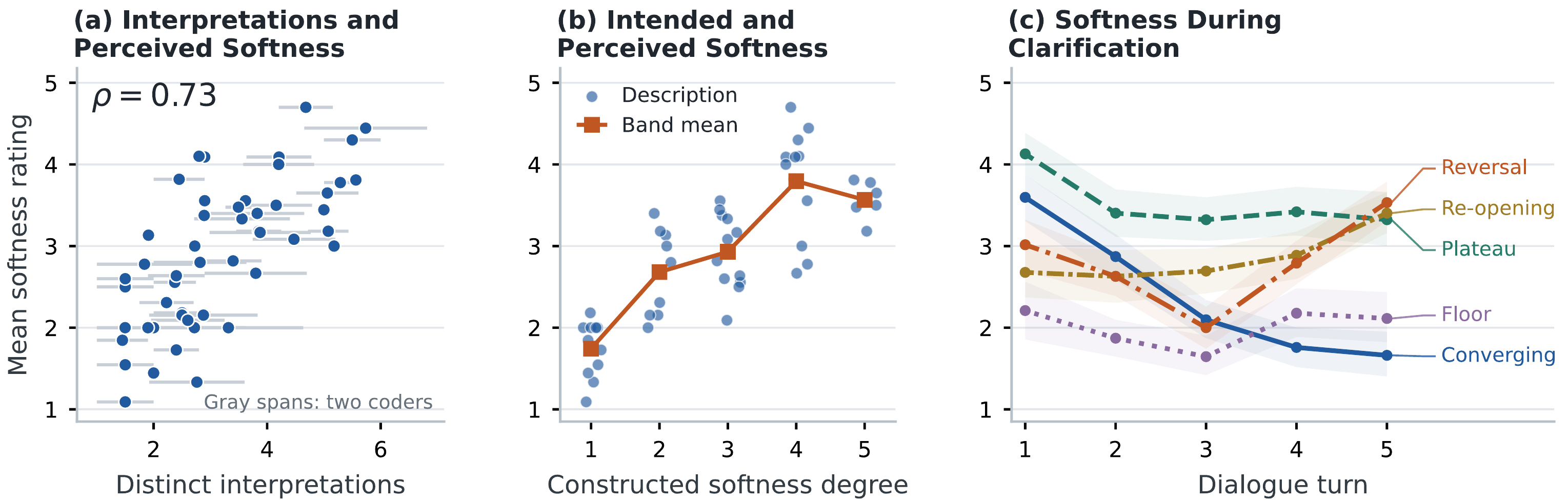}
  \caption{\textbf{Formative Study Results:} (a) Mean softness ratings increase with interpretation diversity as points average two coders’ estimates. (b) Ratings broadly follow constructed softness degrees, with limited separation between degrees 4 and 5. (c) Curves show mean ratings across clarification turns for all five dialogue classes.}
  \label{fig:formative-study-results}
  \Description{Results from the formative study across three analyses, all using mean softness ratings from 1 to 5. (a) Interpretations and Perceived Softness: The 50 descriptions are plotted according to the number of distinct interpretations they received and their perceived softness. Descriptions with more possible interpretations tended to receive higher softness ratings. (b) Intended and Perceived Softness: Human ratings generally increased as the constructed softness degree increased. Mean ratings rose from about 1.8 at degree 1 to roughly 3.8 at degree 4, although degree 5 dropped slightly to around 3.6.(c) Softness During Clarification: Softness changed differently depending on the dialogue type. Converging dialogues steadily became less soft, while Floor dialogues stayed consistently low. Plateau dialogues decreased at first and then remained stable. Re-opening dialogues became softer after a new aspect was introduced, while Reversal dialogues first became more specific and then became softer again after the user changed direction.}
  \vspace{0pt}
\end{figure}

\smallparagraph{Softer Descriptions Elicited More Distinct Interpretations}
Overall, descriptions perceived as softer were also interpreted in more diverse ways (\autoref{fig:formative-study-results}a).
Across the 50 descriptions, the mean rarefied interpretation count correlated with the mean softness rating at $\rho = 0.73$. 
To reduce potential priming and carryover effects, participants provided interpretations before learning our definition of softness, and no participant both interpreted and rated the same description.\looseness=-1

Determining the exact boundaries between readings was less consistent than estimating their number.
Grouping decisions were complicated by interpretations that combined multiple readings and by uncertainty about when additional detail warranted a separate group.
The coders' groupings had a mean Adjusted Rand Index of 0.24, indicating substantial disagreement about exactly which responses expressed the same interpretation. 
Their counts nevertheless differed by at most one for 70\% of descriptions, and the reported reliability of the mean count was 0.79. 
We therefore treat the counts as estimates of interpretive multiplicity, while acknowledging uncertainty in the exact underlying groupings.

These findings support the proposed connection between softness and multiple interpretations within our study setting. 
They motivate estimating softness through interpretive diversity, while leaving open the precise boundaries of the interpretation space.


\newpage
\smallparagraph{Aggregate Judgments Aligned Despite Individual Disagreement}
Mean softness ratings ranged from 1.09 to 4.70, spanning nearly the entire provided scale, and broadly followed the constructed softness degrees ($\rho=.81$) (\autoref{fig:formative-study-results}b).
Although agreement on individual ratings was limited (Krippendorff's ordinal $\alpha=.34$), mean ratings closely aligned with the pairwise comparison scores from the ranking tasks ($\rho=.91$).
Aggregated comparative judgments aligned slightly more closely with the constructed softness degrees ($\rho=.85$) than mean absolute ratings did ($\rho=.81$), suggesting an advantage for assessing softness comparatively.
These findings suggest that the two \textit{judgment tasks} (i.e., rating and ranking) shared structure in relative softness despite variation in absolute judgments.
However, descriptions constructed for the two highest bands (4 and 5) did not separate as intended in either task, limiting support for treating them as distinct categories.
Overall, these results support using aggregated judgments as an empirical reference for relative softness.

\smallparagraph{Clarification Changed Perceived Softness in Both Directions}
In the \textit{clarification task}, softness judgments changed according to the information added during an exchange (\autoref{fig:formative-study-results}c).
Converging dialogues decreased in softness by an average of 1.94 points on our five-point scale from the first to the final turn.
Dialogues that introduced a new facet increased by 0.73 points, while those containing a reversal increased by 0.52 points.

Reversal dialogues showed how later information could undo the effect of earlier clarification.
Their mean rating initially fell from 3.02 to 2.00 as the description became more specific, then rose to 3.53 after a contradictory statement.
Thus, an exchange could become softer again even after an initial reduction.

Plateau dialogues decreased by 0.81 points overall, with the decline occurring at the first clarification and ratings remaining approximately level from the second turn onward.
Floor dialogues remained broadly stable, with a net change of -0.10 points, although ratings dipped slightly before rising again.
Together, these patterns motivate a measure that responds to the content of an exchange and accommodates decreases, stability, and increases in softness.


\subsection{Softness Algorithm}
\label{sec:softness-algo} 

The formative study findings informs our modeling approach in three ways. 
First, the association between possible readings and perceived softness supports interpretive diversity as a basis for estimation. 
Second, the alignment between aggregated ratings and comparisons provides a human reference for relative softness.
Third, the dialogue results motivate a model in which softness can decrease, remain stable, or increase depending on the information added.

We propose an algorithm that probes the latent interpretation space by generating operational readings and measuring their semantic diversity.
The underlying hypothesis is that the separability of a relatively small sample of interpretations reflects the diversity of this broader space, providing an estimate of a description's softness, in line with work that uses the diversity of sampled model outputs as a signal of uncertainty or ambiguity~\cite{kuhn_2023_semantic, cole_2023_selectively}.


\smallparagraph{Approach}
Given an attribute description, a language model generates ten short operational readings, each naming a decision procedure a reader could apply to an essay. 
For example, \emph{well structured} may refer to \emph{a topic sentence in each paragraph} or \emph{logical paragraph ordering}. 
We use high-temperature generation to encourage varied candidates, then embed them with a sentence encoder.
We quantify their diversity using the Vendi score~\citep{friedman_2023_vendi}.
The score represents an effective number of distinct readings, ranging from one for identical embeddings to ten for mutually orthogonal embeddings. 
Unlike counting the ten generated candidates, it accounts for their similarity, assigning lower diversity to paraphrases of the same reading. 
A clipped linear transformation maps this value to a softness score between 0 and 100.

For clarification dialogues, we first restate the transcript as a single current attribute specification, retaining key points while giving precedence to the latest revisions.
We generate readings afresh from this specification after each turn, allowing estimated softness to increase when an exchange introduces new interpretive possibilities.
This restatement step focuses interpretation on the user's expressed intent and reduces the influence of incidental wording in the dialogue.

\begin{figure}[h]
  \centering
  \includegraphics[width=1\linewidth]{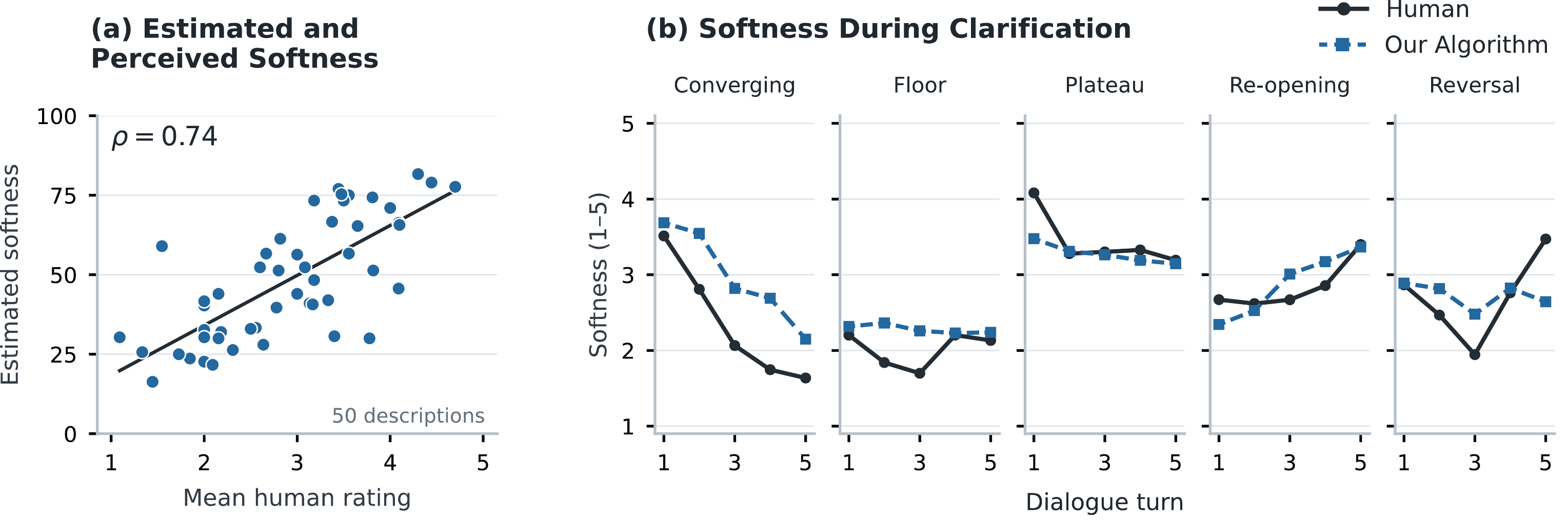}
  \caption{\textbf{Softness Algorithm Results:} (a) Mean softness estimates versus mean human ratings for 50 descriptions. (b) Human and algorithm trajectories across all five clarification classes. Algorithm scores (0–100) are linearly rescaled to the 1–5 human scale for comparison.\looseness=-1}
  \label{fig:softness-measure}
  \Description{A comparison between the computational softness estimate and human softness ratings. (a) Estimated and Perceived Softness: The computational estimate closely follows the mean human ratings across the 50 descriptions, with a strong positive correlation (Spearman ρ = 0.74). (b) Softness During Clarification: The five small plots compare human ratings with the algorithm’s estimates across dialogue turns. For Converging dialogues, both decrease, although the algorithm’s decline is less steep. Floor dialogues remain low for both. In Plateau dialogues, human ratings fall after the first turn and then level out, while the algorithm stays relatively stable throughout. Both humans and the algorithm detect the increase in Re-opening dialogues. For Reversal dialogues, however, humans show a stronger drop and subsequent increase than the algorithm.}
  \vspace{0pt}
\end{figure}

\smallparagraph{Agreement with human observations}
We applied our softness algorithm to each description three times and averaged the resulting estimates.
The intraclass correlation coefficient was $0.887$ for a single run and $0.959$ for the three-run mean.
Across the 50 descriptions, mean estimated softness correlated with mean human ratings at $\rho=0.742$ (\autoref{fig:softness-measure}a).\looseness=-1

The measure also aligned with changes in human softness ratings across turns in clarification dialogues (\autoref{fig:softness-measure}b).
For 100 turns of 20 dialogues, estimated and human-rated softness correlated at $\rho=0.642$. 
After subtracting each dialogue's mean from its respective estimated and human scores, the correlation remained $\rho=0.547$, indicating an association within dialogues as well as between them. 
These results support the measure as an approximation of perceived softness in our study setting.
The estimates remain dependent on the readings generated and the semantic distinctions captured by the sentence encoder.

\section{\app}

In this section, we introduce \app, our tool for interactive soft attribute modeling.
We first describe how its workflow addresses the requirements outlined earlier, then present the agent framework that supports this workflow, and finally introduce the visual, chat-based interface through which users construct and refine attributes.

\subsection{Workflow}

\begin{figure}[t]
  \centering
  \includegraphics[width=\linewidth]{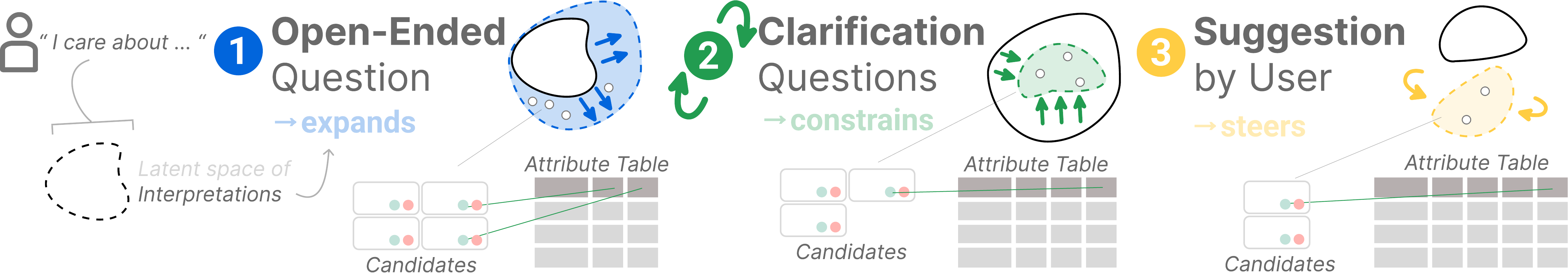}
  \caption{\textbf{Workflow Overview:} (1) Open-ended questions \textit{expand} the space of possible interpretations (2) Iterative clarification questions \textit{constrain} this space. (3) User suggestions \textit{steer} the exploration toward alternative interpretations. Throughout the workflow, candidate attributes are generated and accepted ones end up stored in the attribute table.}
  \label{fig:workflow}
  \Description{A three-stage view of how AttriFy helps users move from a broad idea to useful attributes. The process begins with a user statement such as ``I care about …,'' which may have many possible interpretations. In the first stage, Open-Ended Questions expand the space by encouraging the user to express additional possibilities. In the second stage, Clarification Questions narrow this space by making the intended meaning more specific. In the third stage, Suggestions by the User steer the interpretation toward new aspects the user wants to explore. At each stage, candidate attribute functions can be proposed, accepted, or rejected, and accepted attributes are gradually added to the attribute table.}
  \vspace{0pt}
\end{figure}

We developed an iterative workflow to address the requirements outlined above.
As illustrated in~\autoref{fig:workflow}, our system organizes the interaction around an iterative \textit{\icon{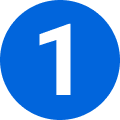} \textbf{expand} $\rightarrow$ \icon{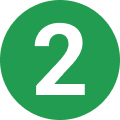} \textbf{constrain} $\rightarrow$ \icon{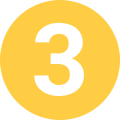} \textbf{steer}} loop that progressively refines attribute definitions to better align them with the user’s mental model.

At the beginning of the interaction, users often hold only vague intuitions or tacit knowledge about the properties they care about. 
Consequently, their initial descriptions are frequently underspecified and admit multiple plausible interpretations – they are \textbf{soft}. 
Following our definition, these interpretations form a latent space of attribute candidates.
Users encounter this space through candidate attributes proposed by the system, while other possible interpretations remain implicit.

Users evaluate these proposals and can accept or reject each attribute, determining which interpretations are operationalized.
Accepted attributes are then computed for individual items and added to an evolving \textit{attribute table}, which represents the current set of confirmed, user-approved attributes and their computed values over the dataset.
The table thus serves as a persistent and interpretable representation of the constructed attribute space.

In the \icon{icons/icon-1.png} \textbf{\textit{expand}} phase, the system surfaces exploratory prompts in the form of \textit{open-ended questions} that help users broaden the space of possible interpretations. 
These prompts support users who may struggle to articulate additional dimensions on the spot, while leaving them free to determine which aspects matter (R2).
This exploration can reveal attribute candidates that were absent from the user's initial description.
To help users move beyond a narrow subset of ideas, the system draws on the current attribute table and the dataset to encourage consideration of complementary dimensions that could distinguish between items (R3), grounding the exploration in the dataset's domain and content.
For example, the system may ask: \textit{``What aspects of these research papers make them feel novel to you?''}

In the \icon{icons/icon-2.png} \textbf{\textit{constrain}} phase, the system asks \textit{clarification questions} that reduce the number of plausible interpretations of the attribute description.
Abstract concepts are decomposed into more concrete aspects or additional constraints are introduced, progressively narrowing the space of attribute candidates and making them more amenable to operationalization.
Specifically, by reducing softness, these clarifications limit the assumptions the system must introduce when approximating the user's intended meaning with a computational function (R4).
This step can be repeated as users progressively articulate and refine their intent (R1).
For example, the system may ask: \textit{``Do you mean novelty in terms of methodology, application domain, or empirical findings?''}

Finally, the \icon{icons/icon-3.png} \textbf{\textit{steer}} phase allows users to provide their own input through \textit{user suggestions}, introducing new directions or contextual information (R6).
Users may also draw on the computed attribute values to identify mismatches between the system's interpretation and their intended meaning.
This mechanism preserves user agency and enables users to correct or redirect the system in response to such mismatches or as their interests evolve.
Depending on the input, steering may broaden the space of interpretations or narrow it further.
For example, a user might specify: \textit{``I am particularly interested in papers that propose simple but elegant ideas.''} 

\subsection{Agent Framework}

\begin{figure}[t]
  \centering
  \includegraphics[width=\linewidth]{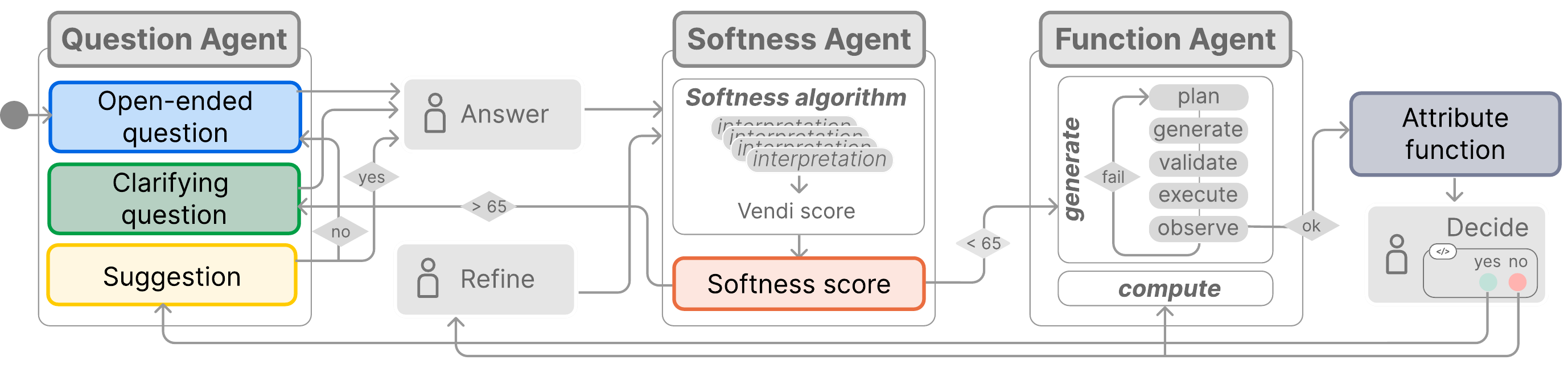}
  \caption{\textbf{Agent Architecture:} The \textit{Question Agent} elicits user intent through open-ended questions, clarification, and suggestions. The\textit{ Softness Agent} estimates interpretive ambiguity and determines whether to continue clarification or invoke the function generation. The \textit{Function Agent} iteratively generates, validates, and tests candidate attribute functions for user review.}
  \label{fig:agent-framework}
  \Description{The system architecture of AttriFy, organized around three agents. The Question Agent manages three types of interaction: open-ended questions, clarification questions, and suggestions. User answers or refinements are passed to the Softness Agent, which samples possible interpretations, calculates a Vendi score, and produces a softness score. If the score is above 65, the system continues asking clarification questions. If it falls below 65, the request is sent to the Function Agent, which follows a repeated plan–generate–validate–execute–observe cycle until it successfully produces an attribute function. The user can then accept or reject the proposed function. Accepted functions are applied to the dataset, while later refinements can send the request back through the Softness Agent. The system can also return to the Question Agent to invite new suggestions.}
  \vspace{0pt}
\end{figure}

\app implements three main types of agents to support our proposed workflow (\autoref{fig:agent-framework}): the \textit{Softness}, \textit{Question} and \textit{Function} agents.
A supplementary technical report with full prompt specifications is provided in Appendix~\ref{app:technical-report}.

\smallparagraph{Softness Agent} calculates a softness score after each user interaction by applying our softness algorithm (\autoref{sec}) to the current discussion.
The score determines whether to invoke the \textit{Function Agent} and which question type the \textit{Question Agent} should generate next.
The thresholds were calibrated using empirical softness evidence collected during the formative study (Appendix~\ref{app:calibration}).
A score of 65 approximately corresponds to the estimated softness of descriptions rated \textit{very soft} or above (mean rating $\geq$4 on our scale), and 30 to descriptions rated \textit{not soft} (mean rating 1 on our scale).
Scores of 65 or above therefore trigger clarification questions without function generation, as selecting among the remaining interpretations would require too many assumptions about the user's intended meaning.
Scores above 30 and below 65 trigger function generation, followed by clarification once users have reviewed the proposals.
Finally, scores of 30 or lower also trigger function generation, followed by an invitation to contribute a user suggestion or begin an entirely new topic.

\smallparagraph{Question Agent} generates different types of questions using the discussion context and previously proposed attributes.
For open-ended questions, it draws on the distribution of items under the current attributes and targets the largest cluster of similarly represented items. 
It uses this cluster to formulate questions that reveal potentially distinguishing properties without referring to individual items by name.
For clarification questions, the agent draws on previously proposed attributes to identify alternative interpretations of the user's description and asks for context or constraints that help differentiate among them.
Prompts for open suggestions invite users to supply directions or contextual information not captured by the preceding questions.

\smallparagraph{Function Agent} uses the CodeAct library\footnote{\url{https://github.com/langchain-ai/langgraph-codeact}} to translate the current discussion block's question--answer pairs into candidate attribute functions.
It generates Python code for individual numerical attributes, drawing on the NLP and embedding libraries available in its execution environment.
Each candidate function is parsed into an abstract syntax tree (AST); candidates that pass validation are executed in a resource-limited sandbox on sample text.
On failure, the agent iterates through planning, generating, validating, executing, and observing until the function works or retries are exhausted.
Functions that pass these checks are stored in a central registry and presented to users through descriptions of their computation. 
Users can accept proposals for dataset-wide computation or refine their intent to guide further generation.\looseness=-1

\subsection{Visual Interface}

\begin{figure}[t]
  \centering
  \includegraphics[width=1\linewidth]{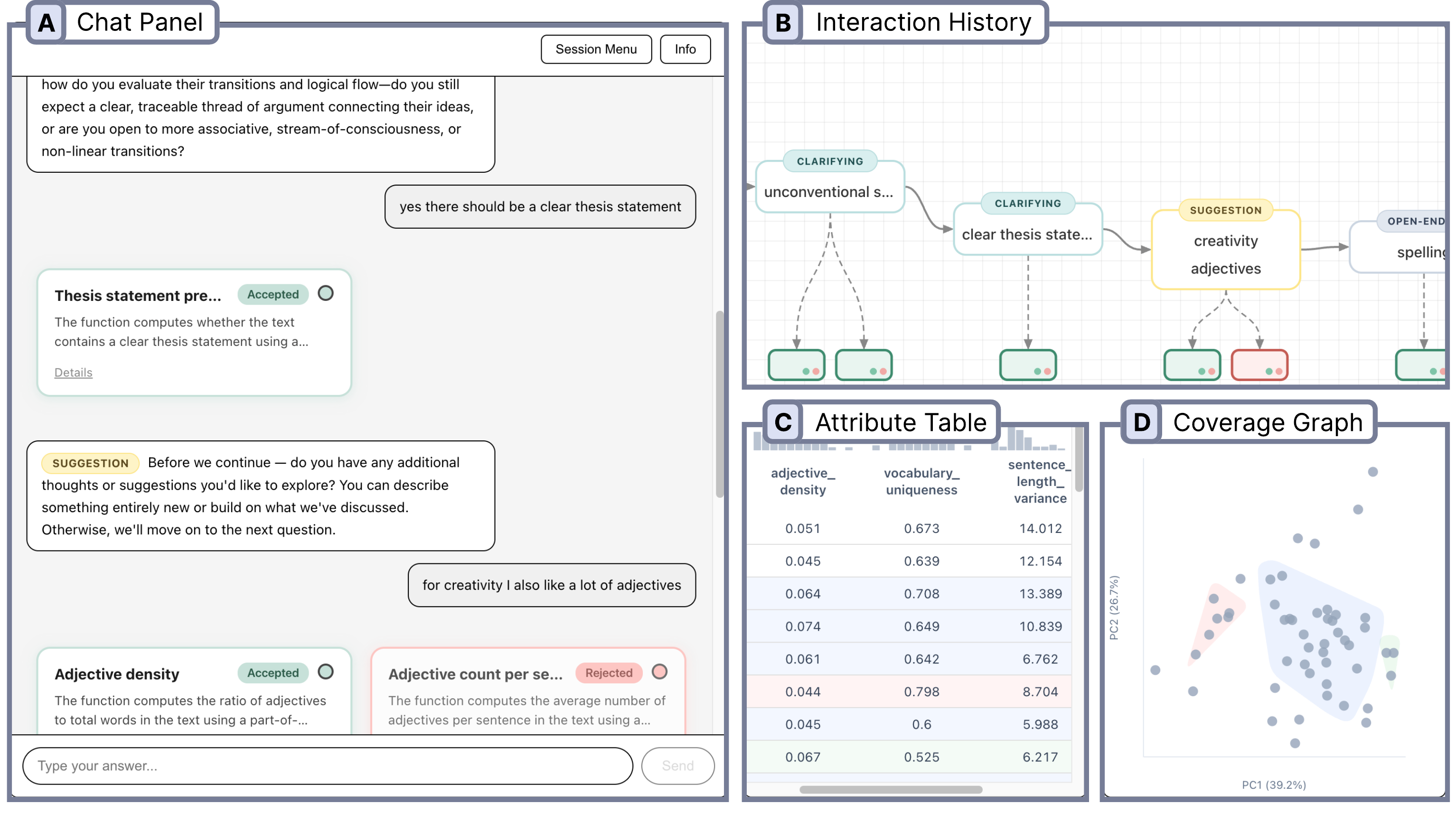}
  \caption{\textbf{The \app interface:} (A) \textit{Chat Panel} supports natural language interaction, and review of candidate attribute functions. (B) \textit{Interaction History} provides visual provenance by linking discussion steps to accepted and rejected candidates. (C) \textit{Attribute Table} displays computed attribute values for dataset items. (D) \textit{Coverage Graph} projects items into two dimensions based on their attributes, showing clusters of similarly represented items.}
  \label{fig:interface}
  \Description{An overview of the AttriFy interface, divided into four main panels. (A) Chat Panel: The user and system discuss essay preferences. For example, the user specifies that essays should have a clear thesis statement, leading to an accepted thesis-related attribute. Later, the user says that creative writing should use many adjectives. AttriFy proposes two related attributes, ``Adjective density'' and ``Adjective count per se…,'' of which the first is accepted and the second rejected. (B) Interaction History: A visual timeline shows previous clarification, suggestion, and open-ended interactions, along with the attribute functions generated from them. Rejected functions are visually marked. (C) Attribute Table: Accepted attributes, such as adjective density, vocabulary uniqueness, and sentence-length variance, are shown as columns containing a numeric value for each essay. Small histograms summarize the distribution of each attribute. (D) Coverage Graph: A two-dimensional scatter plot shows how essays are distributed across the resulting attribute space, including major clusters and outliers.}
  \vspace{0pt}
\end{figure}

\app uses a visual interface to guide the user through our workflow.
The interface comprises four coordinated panels, shown in~\autoref{fig:interface}: (A) the chat panel, (B) interaction history, (C) attribute table, and (D) coverage graph.

\smallparagraph{Chat Panel}
The chat panel supports natural-language conversation through a sequential question-and-answer exchange, presenting one question at a time.
When the softness score falls below the target threshold, the chat reveals the possible function definition as a card containing its name and a detailed description of how it will compute the attribute values.
This description allows users to assess what the function actually measures, since a broad name such as \texttt{calculate\_creativity} may conceal a simple counting algorithm (R5).
Users must accept, reject, or request refinement of each proposed function before the system asks the next question (R6).
This sequence keeps the interaction focused on the current question and its associated proposals.

\smallparagraph{Interaction History}
The interaction history presents a visual timeline of the session, with each interaction positioned vertically according to its softness score.
Nodes are labeled with keywords from the user's response and indicate the question type. 
Expanding a node reveals the full corresponding question and answer.
Functions proposed at an interaction appear as linked cards similar to those in the chat and indicate whether they were accepted, rejected, or refined.
These cards are placed at the zero-softness baseline to represent a single computational interpretation. 
The vertical distance from an interaction node to this baseline visually represents the interpretive gap (or approximation gap) bridged by the computational approximation, with greater distances indicating that more possible interpretations remained when the function was generated.
Users can revisit earlier cards and revise their decisions, for example when a newly generated function makes a previously accepted attribute redundant (R6).
The history thus  traces how softness changes and which functions emerge over the course of the interactive development of the attribute space (R5).

\smallparagraph{Attribute Table}
When a user accepts a function card, the system applies the function to the dataset and displays the resulting attribute values in the table, connecting the natural-language dialogue to its computational results.
Users can inspect these values alongside the underlying items through the same panel, allowing them to assess how well the functions capture their intended meanings.
Histograms above the attribute columns summarize the distribution of values across the corpus.

\smallparagraph{Coverage Graph}
The coverage graph visualizes the separability of items under the current attributes (R3).
Items are projected from the attribute table into two dimensions and displayed as points, with clusters highlighted to reveal groups that are similarly represented by the constructed attributes.
As the attribute table changes, the projection updates to show how the representation of the corpus evolves.
This view helps users identify groups of items that may warrant additional distinguishing attributes.
The same clusters guide the Question Agent's open-ended questions, which target the largest cluster to elicit attributes that could further distinguish its items.

\subsection{Illustrative Use Case}

Consider a teacher who wants to define grading criteria (i.e., attributes) for argumentative essays. 
The teacher tells \app that an essay should ``take a clear stance.'' 
In this example, the \textit{Softness Agent} estimates a softness score of 58 and the \textit{Function Agent} proposes a computational approximation for review while also inviting further clarification.
A proposed function might detect a thesis-like sentence in the first paragraph. 
Its card describes this procedure, allowing the teacher to consider whether it captures an aspect of the intended criterion.
Accepting the proposal adds its computed values to the attribute table, where the teacher can inspect the distribution and compare values with individual essays.

After proposal review, the \textit{Question Agent} asks whether the stance must also be maintained throughout the essay.
The teacher responds that the position should not change. 
This answer adds a constraint to the current specification, which the system reassesses before selecting its next action. 
If the estimated softness is sufficiently low, the system proposes functions for review and subsequently invites the teacher to suggest another aspect or begin a new topic.

The interaction history links each proposal to the discussion that produced it. 
Function cards appear at the zero-softness baseline to indicate a selected computational interpretation.
Together, the history and attribute table allow the teacher to trace proposed interpretations and inspect their consequences. 
This example illustrates the workflow rather than reporting a recorded study session.


\subsection{Implementation Details}
\app is implemented as a browser-based workspace supported by a Python service.
The client is a React and TypeScript single-page application that presents the four components described above.
The server is a FastAPI application that stores session state in MongoDB.
Dialogue generation and softness scoring use Google Gemini, with \texttt{gemini-3.5-flash} for questions and restatement and \texttt{gemini-3.5-flash-lite} for sampling interpretations.
These model choices reflect the different latency requirements of the two tasks.
The softness algorithm combines sentence-transformer embeddings with a Vendi-derived score.
Natural-language attribute descriptions are compiled into executable Python scoring functions by an external CodeAct service (\texttt{langgraph-codeact}).
Accepted functions are applied to the essay collection, after which scikit-learn PCA and DBSCAN produce the clustered projection.

\section{User Study}

We conduct a controlled, web-based within-subjects study comparing \app with an LLM-driven baseline.
The evaluation examines support for attribute construction, allowing participants to develop and revise their intended specifications during the interaction.
We address three research questions:

\begin{itemize}[leftmargin=23pt,labelsep=0.3em]
    \item[\textbf{RQ1:}] \textbf{Clarifying intent -} How does \app affect participants' perceived ability to express and clarify their intended attribute specifications?
    \item[\textbf{RQ2:}] \textbf{Interaction experience -} How does \app affect participants' perceived effort, perceived transparency, control and ownership over attribute construction?
    \item[\textbf{RQ3:}] \textbf{Interaction patterns -} How do interaction patterns, including attribute generation and revision patterns differ between \app and the baseline?
\end{itemize}

\subsection{Participants and Study Design} 

We recruited 41 participants through Prolific, restricting eligibility to native English speakers aged 18 or older. Of these, 32 passed the attention check and completed the study. 
We refer to these participants as P1--P32 throughout the paper. 
We report participant demographics, prior experience with LLM tools, and exclusion criteria in the Appendix~\ref{app:user-study-details}.

Each participant completed two blocks in a single session, one with \app and one with the Baseline.
Tool order and dataset assignment were counterbalanced across four allocation cells, crossing the starting tool (\app or baseline) with the starting essay dataset (argumentative or narrative). 
Each participant saw different essay materials in each condition, so that tool comparisons were not confounded by familiarity with specific texts.
The complete session lasted approximately 60-90 minutes, and participants received compensation at a rate of 12\$ per hour.

The study was approved by our institution’s institutional review board (details omitted for anonymous review).
All participants provided written informed consent prior to participation.

\subsection{Baseline, Tasks, and Procedure}

\smallparagraph{Baseline}
The Baseline provides an LLM chat interface with access to the same function-generation service as \app.
To ensure a fair isolated comparison, we also show the participants the same attribute table (\autoref{fig:interface}c) and projected coverage graph (\autoref{fig:interface}d).
However, its conversation is not governed by our tool's structured workflow or softness scores.
Where \app uses estimated softness to determine when to clarify and when to generate functions, the Baseline generates functions at every step.
Participants still explicitly accept, reject or refine proposed functions through the chat.
This focuses the comparison and evaluates the complete \app workflow against an open-ended LLM driven conversation, with both workflows having access to the same generation tools and supporting panels.

\smallparagraph{Tasks}
Participants constructed attributes for collections of approximately 500-word essays, using a narrative essay dataset in one block and an argumentative essay dataset in the other.
All the data used in the study is sourced from the ASAP 2.0 corpus comprising of 24,278 student essays\footnote{https://www.kaggle.com/datasets/lburleigh/asap-2-0}.

\smallparagraph{Procedure}
Each session lasted approximately 60-90 minutes. 
After consent and a brief introduction, participants completed a general tutorial using materials separate from the main tasks.
Before each study block, they received a description of the assigned dataset and a tutorial specific to the tool used in that block.
Participants then accessed the assigned tool and had 15 minutes to construct a set of attributes reflecting the properties they considered important in the assigned essay collection. 
A visible timer indicated the remaining time.
The system initiated the conversation, after which participants could freely interact with the tool, discuss their ideas, and accept, reject, or refine proposed functions.
At the end of each block, participants reviewed their resulting specifications and completed a questionnaire and written reflection.
After both blocks, they indicated their comparative preference and explained their choice.
Further procedural details are provided in Appendix~\ref{app:user-study-procedure}.

\subsection{Measures and Analysis}

To comprehensively address our research questions, we employed a combination of self-reported and behavioral measures, complemented by qualitative analyses of participants’ free-text responses.

\smallparagraph{RQ1 \& RQ2: Clarifying intent and interaction experience}
We assessed participants' perceptions of each tool using a customized questionnaire with 10 items rated on a 5-point Likert scale (1 = strongly disagree, 5 = strongly agree). 
The items were developed to target specific aspects of the interaction experience relevant to our research questions, including perceived ability to clarify intent, as well as perceived transparency of outputs, and control and ownership over the resulting attributes. 
In addition, we used the NASA-TLX~\cite{hart_1988_development} to measure participants perceived effort and overall workload. 
The complete questionnaire is provided in Appendix~\ref{app:user-study-questionnaire}.

\smallparagraph{RQ3: Interaction patterns}
To measure participant's interaction behavior, we record session duration, system waiting time, conversational turns, generated functions, approvals, rejections, and revisions.
These records characterize the construction process and fewer generations or higher acceptance are not treated as evidence of better attributes.

Finally, after each condition, participants described a specific instance in which the interaction clarified, changed, or failed to capture their clarification process.
The closing questionnaire asked which tool they would choose for a similar task, including an option for no preference, and why.

\smallparagraph{Analysis}
As our study employed a within-subject design and Likert-scale questions, we used Wilcoxon signed-rank tests~\cite{wilcoxon_1945_individual} with Holm–Bonferroni correction for multiple comparisons, to compare \app against the Baseline.
We further conducted a qualitative review of participants' free-text responses to identify recurring observations and provide additional context for the quantitative findings.

\subsection{Results}

\subsubsection{RQ1: Clarifying intent.}
\begin{figure}[t]
  \centering
  \includegraphics[width=1\linewidth]{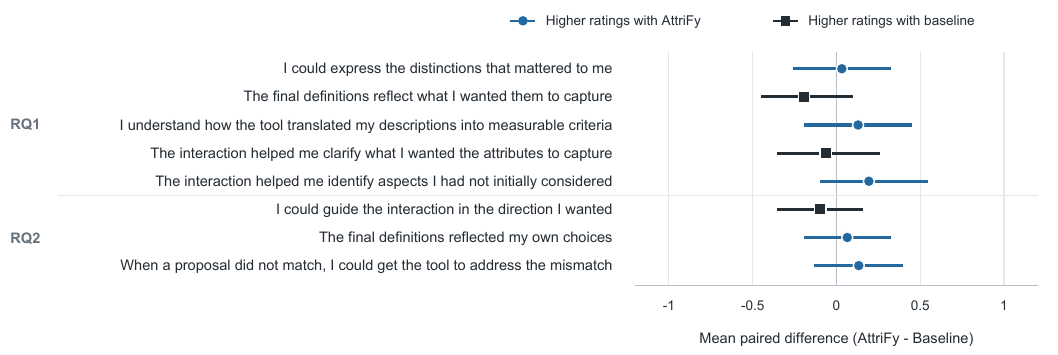}
  \caption{Mean paired differences in questionnaire ratings between \app and Baseline. 
  Points show the participant-level mean paired difference (\app-Baseline), and horizontal lines show 95\% confidence intervals. 
  Positive values indicate higher ratings under \app and negative values indicate higher ratings under Baseline. 
  Ratings were collected on a 1-5 Likert scale. }
  \label{fig:diffs_forest_plot}
  \Description{Differences between AttriFy and the Baseline across eight Likert-scale questionnaire items. Positive values indicate higher ratings for AttriFy, while negative values favor the Baseline. For RQ1, AttriFy scores slightly higher on expressing important distinctions (+0.03), understanding how descriptions were translated into measurable criteria (+0.13), and identifying aspects not initially considered (+0.19). The Baseline scores slightly higher on whether the final definitions captured the intended meaning (−0.19) and whether the interaction helped clarify the desired attributes (−0.06). For RQ2, the Baseline scores slightly higher on being able to guide the interaction (−0.09), while AttriFy scores slightly higher on whether the final definitions reflected the user’s choices (+0.07) and whether mismatched proposals could be corrected (+0.14). All 95\% confidence intervals cross zero, meaning none of these differences is statistically reliable.}
  \vspace{0pt}
\end{figure}

Both \app and the Baseline received high overall ratings with respect to participants perceived ability to express and specify their intended attributes ($M = 4.13-4.58$).
These ratings suggest that participants perceived both tools as supporting their ability to generate and clarify attributes. 
No statistically significant differences were found between \app and Baseline on the questionnaire items for expression of intent and development of ideas after Holm correction ($p>.05$ for all comparisons). 
However, the point estimates showed descriptive differences in both directions. 
\app received higher ratings for identifying previously unconsidered attributes and understanding how attributes were translated into items, whereas the Baseline received slightly higher ratings for alignment of the final definitions (see \autoref{fig:diffs_forest_plot}).  

To understand what these similarly high ratings reflected in practice, we examined the users' free-text feedback after each session. 
Both tools allowed users to reflect and clarify their intent. 
P1 reported that \app ``helped when it asked me to think about the difference between minor grammar mistakes and errors that actually affect clarity.'' P17 in describing the Baseline reported:``When it broke \textit{voice} into small pieces like sensory details and vulnerability it finally clicked for me.`` In both cases, the tool's questions prompted participants to decompose an intuitive judgment into components they could articulate.

Several participants attributed their ability to clarify intent to the tools' questioning itself. 
P24 noted: ``When I said what I looked for in a narrative essay was clarity, \app had to ask if that was all what I look for, that question made me carefully search within me, remembering those attributes that makes a narrative essay captivating and useful to me.'' and further, ``... at the moment I was not sure, but the question provoked my thought process and brought out my real authentic reflection.''
However, some clarification attempts inadvertently hindered participants' ability to express their intended attributes.
P11 reported that ``\app hindered my developing of attributes when it got stuck on my answer to whether I preferred ambiguity or leaving no loose ends. 
It could not move past this issue ... which wasn't something I was interested in at all.'' 
In other words, \app's attempts at a computable attribute hampered the participant's expressed preference, rather than developing it.  
Interestingly, P16 reported:``\app kept asking me to clarify, essentially, how to tell that an essay is actually good. I felt like I was coming up against its fundamental absence of taste.'' 
This raises the question of whether such elicitation can accommodate attributes that participants recognize intuitively but resist operationalization.

When asked which system they would choose for a similar task, 12 participants chose \app, 16 the baseline and 4 had no preference (binomial, $p = .57$).
Order explains the response better than system, as 20 of the 28 with a preference chose whichever system they used second ($p = .036$).
We therefore do not treat this item as a clear measure of system preference. 
Because participants used a different system for each essay genre, their explanations could also reflect the materials: 14 reasons focused on essay content, while the 14 interaction-related reasons (e.g., asking more follow-up questions) referred equally often to \app and the baseline (7 each).

\subsubsection{RQ2: Interaction experience}
Participants reported high perceived control and ownership in both conditions, with means between 4.20 and 4.42 on the three control and ownership items (\autoref{fig:diffs_forest_plot}, lower half). 
No comparison reached significance after Holm correction. 
When asked whether participants felt like they could address the mismatch between proposed and intended attributes effectively, 16 participants selected ``not applicable," indicating they did not perceive a mismatch to correct. 
Some friction in perceived control was reported by P17: ``\app kept pulling me back to facts and structure which isn't how I know a piece is honest''. 
This may indicate that clarification can come across as too restrictive for some users. 

NASA-TLX subscale ratings were similar across conditions, including perceived effort, with no subscale differing by more than 6.1 points on the 0--100 scale ($M = 43.3$ vs.\ $M = 43.1$; all $p > .28$). 

\subsubsection{RQ3: Interaction patterns}

In addition to participants' subjective experience, we analyzed the details of their attribute construction process. 
\app sessions involved significantly fewer function generation events ($M = 5.2$, $SD = 2.6$) than baseline sessions ($M = 8.7$, $SD = 4.2$; $p < .001$). 
This pattern is consistent with \app's softness-based generation criterion, which reserves function generation for descriptions assessed as sufficiently specified and otherwise continues asking questions. 
The baseline generated new attribute sets without this criterion, averaging 8.7 generation events per session. 
In contrast, \app combined initial generation ($M = 2.4$) with refinement following further clarification ($M = 2.8$). 
This reflects the intended workflow of gathering sufficient information before generation and subsequently updating the operationalization as the description develops.

\begin{figure}[t]
  \centering
  \includegraphics[width=1\linewidth]{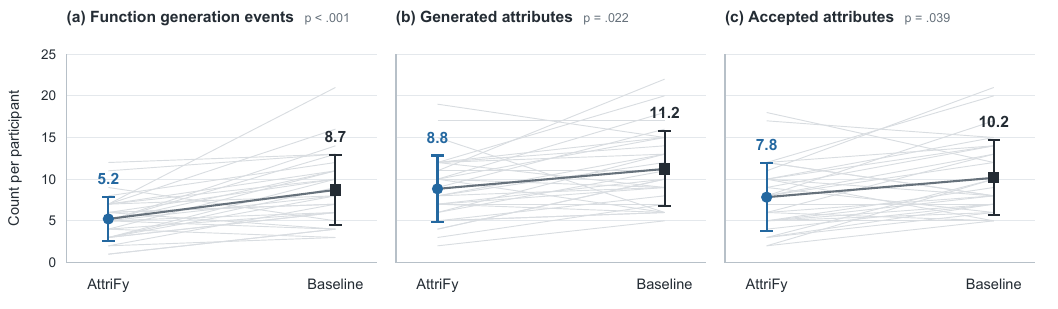}
  \caption{\textbf{Interaction process measures:}(a) Number of generation events, including initial generation and regeneration. Each event can produce multiple attributes. (b) Total number of attributes generated, regardless of acceptance. (c) Number of attributes accepted. Gray lines connect each participant's values across conditions. Markers indicate condition means with $\pm$1 standard deviation.}
  \label{fig:process_measures}
  \Description{A participant-level comparison of how many generations and attributes were produced with AttriFy versus the Baseline. Each participant’s two values are connected by a gray line, while the darker line shows the condition averages. (a) Participants required significantly fewer function generations with AttriFy: 5.2 on average compared with 8.7 for the Baseline (p < .001). (b) AttriFy also resulted in fewer accepted attributes: 7.8 compared with 10.2 (p < .05). (c) The total number of attribute function cards was likewise lower with AttriFy: 8.8 compared with 11.2 (p < .05). Although individual participants varied considerably, the Baseline generally produced more generations and more attributes.}
  \vspace{-1em}
\end{figure}

\app also produced fewer attributes overall ($M = 8.8$, $SD = 4.0$ versus $M = 11.2$, $SD = 4.5$; $p = .022$; \autoref{fig:process_measures}b) and fewer accepted attributes ($M = 7.8$, $SD = 4.1$ versus $M = 10.2$, $SD = 4.5$; $p = .039$; \autoref{fig:process_measures}c). Acceptance rates were similar across conditions, at 88\% for \app and 91\% for the baseline. 
Thus, the difference in accepted attributes accompanied a difference in how many attributes were generated, rather than a substantial difference in the proportion accepted. 
Among accepted attributes, baseline sets were also more redundant: participants' mean nearest-neighbor embedding similarity was higher than under \app ($M = 0.80$ versus $M = 0.77$), with more near-duplicates at cosine $\ge 0.85$ per participant ($M = 2.1$ versus $M = 0.8$).\looseness=-1

To examine the computational behavior of the resulting attributes, we re-executed the constructed functions on a subset of 200 essays. 
Most functions used hardcoded word lists (63\%) or surface statistics (26\%), with fewer using syntactic features (9\%) or named entities (2\%). 
Overall, 38\% of the functions returned zero values for more than half the essays in the sample.
Functions with near-identical descriptions generally ranked essays similarly (median $\rho=.75$, compared with $.08$ for unrelated pairs). 
Within \app, functions generated from higher-softness answers more often returned zero for most essays than those generated from lower-softness answers (40\% versus 26\%).

Analysis of the \app interaction traces showed that recorded softness did not consistently decrease during the dialogue.
Across 184 consecutive response pairs within discussion topics, recorded softness increased in 85 cases (46.2\%), decreased in 89 (48.4\%), and remained unchanged in 10 (5.4\%). 
Of the 67 topics, 30 ended with higher softness than they began and 37 ended with lower softness; the median change was $-2$ points. 
For example, P3 elaborated an initial description of strong reasoning into requirements for empirical evidence, credible peer-reviewed sources, and integration of findings, while recorded softness increased from 39 to 47, 55, and 62.

Taken together, these results show that \app involved fewer generation events and fewer generated attributes, while acceptance rates remained similar. 
Continued clarification did not consistently reduce recorded softness, and higher-softness answers were associated with functions that more frequently returned zero for most essays.

\section{Discussion}

In this chapter we discuss what our two studies reveal about understanding softness and how \app supports users compared to the LLM-baseline.

\subsection{Softness}

Our findings provide empirical support for defining softness as the diversity of valid interpretations a description permits. 
Descriptions that elicited more distinct interpretations by study participants were perceived as softer, and aggregated ratings converged with comparative judgments. 
Together, these results ground softness in observable patterns of interpretation and judgment within the studied setting.

In addition, our results show that participants were more aligned with our constructed softness degrees when making comparative judgements between descriptions than when assigning absolute ratings on our pre-defined scale. 
This suggests that softness may convey a useful signal without requiring a precise numerical value. 
One possible explanation is that people recognize softness or interpretive openness intuitively, but they lack a convention for expressing it explicitly.

The limited separation between the two highest constructed softness degrees suggests that distinctions at the softer end were less clearly resolved. 
This reflects the asymmetry in the scale: low softness can be anchored in a single operational interpretation, whereas high softness reflects an open-ended latent space of possible readings. 
People may envision different portions of that space, assigning different scores to the same description. 
This variation itself reflects a property that a softness measure should capture to work effectively.

Although interpretive diversity aligned with perceived softness overall, identifying the boundaries between interpretations remained difficult. 
Coders agreed more closely on the number of groups than on which responses exactly belonged together. 
Annotating responses that combined multiple readings or provided nuanced additional details involved considerable subjectivity as it was difficult to determine how these differences would be reflected in the selected essays.
The annotations therefore support the presence and approximate extent of interpretive diversity, while leaving its precise structure uncertain. 
This motivates the need for clarification to gain insight into what exactly users mean and which distinctions matter to them.


Our proposed softness measure shows strong alignment with the recorded human judgments, and its design reflects observations from the formative study.
Rather than identifying exact interpretations or fixed clusters, the algorithm estimates how distinct a sampled set of plausible interpretations is. 
Our results suggest that this sampled diversity provides a useful proxy for perceived softness within this setting.




\subsection{Attribute Construction}

Our exploratory user-study yielded more mixed results. 
Within \app, recorded softness increased almost as frequently as it decreased within discussion topics, suggesting that continued conversation does not necessarily produce a steady progression towards an executable attribute. 
One participant (P3), for example, specified increasingly detailed and nuanced requirements for evidence and its integration while softness was shown to increase.
A possible reason is that clarification exposes additional complexity: relationships, exceptions, and contextual distinctions can make a construct more explicit in theory without making it easier to compute.
These additional requirements may admit several plausible implementations, leaving ambiguity about how to combine them into a coherent measurement.
This reflects a more general limitation of free-form verbal question-answering. 
Although the system can identify unresolved aspects and guide further clarification, it cannot ensure that subsequent responses reduce softness. 
Participants may instead introduce additional distinctions that expand the construct and make the operationalization more difficult than before.

We expected the baseline's larger approximation gap to lead participants to accept fewer attributes, as translating less explored descriptions into executable functions would require more assumptions about their intent.
The observed patterns did not support this expectation, raising questions about what acceptance actually tells us about attribute construction.
The execution analysis reinforces this concern as many functions returned zero for the majority of essays.
Although sparsity does not make an attribute invalid, acceptance alone probably cannot establish whether it reflects the intended behavior.
This may stem from reduced user attention during the evaluation of AI-generated content. 
As~\citet{beck_2026_bias} demonstrate, people frequently over-rely on AI suggestions and accept flawed outputs rather than inspecting them independently.
This is particularly the case when verification demands additional effort or when users hold a strong bias toward AI capabilities.
Together, these findings highlight the need to distinguish recognizing an attribute as relevant from assessing whether its implementation measures what was intended.


\section{Limitations \& Future Work}

Our findings highlight opportunities for interactive soft attribute construction, while leaving open questions about functional alignment, the evaluation of attribute sets, implicit signals and integration with preference elicitation. 
We discuss these limitations and directions for future work below.


\smallparagraph{Assessing Functional Alignment}
Our study measured participants' perceptions and decisions, but did not independently assess whether generated functions captured their intended preferences.
A preliminary exploratory inspection revealed discrepancies between participants’ stated preferences and accepted functions, but did not establish how common these discrepancies were or whether participants knowingly accepted them.
Acceptance and revision counts therefore cannot be interpreted as direct measures of functional alignment.
To address this limitation, we plan a follow-up annotation study with our collected data. 
Independent annotators would assess alignment between participants' stated preferences, the generated descriptions, and the underlying implementations, initially without access to condition labels or participants' decisions. 
Comparing these assessments with acceptance and rejection behavior would help assess whether participants accepted functions that aligned with their stated intentions.
However, it should be noted that this analysis alone would not establish how critically participants assessed the outputs. 
Understanding their reasoning would require complementary evidence, such as retrospective interviews.

\smallparagraph{Evaluating Attribute Sets}
Assessing a final attribute set requires considering not only the attributes constructed, but also potentially relevant dimensions that remain unexplored.
Users may not recognize these omissions until prompted by examples or questions, so self-reported satisfaction cannot establish adequate coverage of their interests. 
Behavioral findings are similarly difficult to interpret. 
Continued exploration may indicate discovery or difficulty, while moving on may indicate satisfaction or a desire to minimize effort.
Our measures cannot confidently establish which of these explanations accounts for participants' behavior.
Therefore, future evaluations could look to use more concrete examples and downstream comparison tasks to reveal missing distinctions and examine how users revise their attribute sets. 
For long-form essays, such evaluations would require additional time or focused comparisons of selected items.
Tracking these revisions alongside users' explanations could help distinguish newly articulated interests from changes in what they consider important.

\smallparagraph{Implicit Attribute Construction} 
The potential user over-reliance on AI outputs motivates a shift toward implicit preference inference, where systems infer constraints directly from natural dialogue rather than relying on upfront validation steps. 
In a potential adaptation of our workflow, our interface could passively track the ongoing dialogue to compile and test candidate Python functions in the background. 
By using each conversational turn to refine which functions hold, the system infers underlying preferences implicitly, avoiding real-time cognitive friction or the risk of uncritical acceptance of suggestions.
While prior work has explored capturing feedback through targeted questioning~\citep{li_2025_gate,dou_2026_gate}, structuring free-form dialogue so that routine responses yield reliable functional specifications remains an open challenge. 
Furthermore, as~\citet{su_2026_tell} note, because purely implicit signals can be ambiguous, an effective interface must balance background inference during dialogue with selective, late-stage opportunities for explicit user review—where, in our setting, the system could present the compiled functions at the end of the interaction for final user inspection and feedback.

\smallparagraph{Integrating Preferences}
We scoped this work to attribute construction as a foundation for downstream preference elicitation, establishing the dimensions along which users can express likes and dislikes. 
However, users' descriptions may already convey both an attribute of interest and a preferred direction or value. 
Treating these descriptions solely as input to attribute construction leaves this preference information unused.
Future work could jointly support attribute construction and preference elicitation, allowing users to inspect their combined effects through item rankings or recommendations. 
These outputs could make the consequences of their specifications easier to assess than attribute definitions alone. 
For example, a user may initially favor adjective-rich essays but reconsider after inspecting highly ranked examples. 
Such feedback could motivate revisions to the attribute definition, the preference expressed over it, or both.
This could also help address the attribute set evaluation limitation discussed above.

\section{Conclusion}
Before modeling what users like or dislike, it is necessary to establish the attribute dimensions along which their preferences can be meaningfully expressed.
We introduced \app, a system for interactive construction of soft attributes from textual corpora.
Our approach uses \textbf{softness} to characterize the space of plausible attribute interpretations and guide an iterative workflow of expanding, constraining, and steering. 
By combining conversational guidance, executable attribute functions, and visual inspection, \app enables users to develop their intended meanings and examine how they are translated into computational representations.
We view this work as a step toward tools that support users in shaping the dimensions through which they understand and compare data, while retaining control over how those dimensions are defined and computed.

\section{GenAI Usage Disclosure}

We used AI tools, including Claude Code (Claude Sonnet 5, Claude Opus 4.8/5), during this project. 
The problem framing, study design, modeling, and evaluation were defined and executed by the authors. 
AI tools supported literature exploration, comparison of implementation options, exploratory analysis of the large-scale interaction log data, and feedback on design and writing decisions. 
All AI-suggested sources and claims, as well as any patterns surfaced during exploratory analysis, were independently verified by the authors before use.

As discussed in the main paper Claude Opus 4.8 was used to generate the candidate stimuli for the formative study (attribute descriptions and clarification dialogues), which the authors then manually screened.
For the implementation and evaluation, we developed the initial structure and core content ourselves and used AI assistance to iterate on specific parts under our direction, including smaller utility tasks such as formatting and plotting scripts. 
When AI outputs were used, we specified constraints, reviewed the results critically, and integrated, revised, or rewrote the final material ourselves. 
The central ideas, codebase, argumentation, and conclusions of this work are our own, and the authors take full responsibility for all content in this manuscript.


\newpage
\bibliographystyle{ACM-Reference-Format}
\bibliography{main}

\newpage
\appendix

\section{Softness Study}
\label{app:formative}

\subsection{Softness Scale}
\label{app:softness-scale}
Participants received the definition of softness as the extent to which a
description admits multiple valid interpretations without further
specification. The following anchors accompanied the five-point scale.
\begin{enumerate}
\item \textbf{Not soft}. One interpretation. Anyone applying the description would arrive at the same reading.
\item \textbf{Slightly soft}. One clear interpretation, with minor variation in details that would rarely change the outcome.
\item \textbf{Moderately soft}. A few defensible interpretations, roughly two or three, with one of them somewhat more natural than the rest.
\item \textbf{Very soft}. Several defensible interpretations, none of them clearly dominant.
\item \textbf{Extremely soft}. Many defensible interpretations, spread widely, with no natural starting point.
\end{enumerate}

\subsection{Study Details}
\label{app:study-details}
Participants were recruited through Prolific with eligibility restricted to
English fluency and university-equivalent education. 
The three tasks occurre dn a fixed order. Participants first interpreted ten descriptions, then rated ten descriptions and ranked eight triplets, and finally rated five dialogues
over five successive turns. 
Interpretations were elicited before introducing the softness definition. 
No participant interpreted and individually rated the same description. 
This separation did not extend to the individual-rating and ranking tasks.
Patticipants were compensated at a hourly rate of 12\$ (9£) with the experiment advertised at an estimated duration of 30min.

\subsubsection{Attention Checks and Exclusions}
The analysis retained 62 of 70 recorded sessions after removing seven incomplete sessions and one duplicate. 
All 62 retained sessions passed both attention checks.
Our attention check in the interpretation task asked participants to answer ``yes'' in a particular field, while the ranking check explicitly marked three descriptions as most, medium and least soft and required participants to place them in the corresponding order.  Completeness checks required the expected number of distinct items and valid responses in all
three tasks, including all five ratings for each assigned dialogue.
The retained data comprise 620 interpretations, 620 individual ratings, 496 triplet rankings and 1,550 turn ratings from 310 dialogue viewings.
Across the 50 descriptions, individual ratings numbered 8--22 per description (median 11). Each triplet contributes three pairwise comparisons, giving 1,488 comparisons in total.

\subsection{Stimuli}
\label{app:stimuli}
The pool contained 50 essay-attribute descriptions and 20 five-turn clarification dialogues. 
Candidate stimuli were generated using Claude Opus 4.8 (June 2026), and manually screened by the authors to remove duplicates and implausible user descriptions. 
Constructed softness degrees and dialogue classes express design intentions, rather than human measurements or exact numerical predictions of subsequent ratings.

\subsubsection{Description Generation Prompt}
To construct the softness dataset, we prompted an LLM with the instructions below, specifying the task, three annotation axes (\emph{Domain}, \emph{Source of Softness}, \emph{Degree of Softness}), and the required output fields per entry.

\begin{promptbox}[Dataset Generation Prompt]
\begin{lstlisting}[style=promptcode]
Task: Create a dataset of 50 single, atomic essay attribute descriptions. Each entry must describe exactly one specific attribute (e.g., separate "spelling accuracy" from "punctuation correctness") and map it across three axes: Domain, Source of Softness, and Degree of Softness.
It is important to use language that is accessible to an average reader and the difficulty should not stem from overly complex phrasing.

Core Constraints:
- Ultra-Simple Language: Write all descriptions and explanations in plain, everyday words that anyone can instantly understand, regardless of their background or training. Avoid technical, academic, or textbook terms (e.g., no terms like open scale, threshold, judge parameter, syntactic, disjunctive, or orthographic).
- Single Atomic Attribute: Focus strictly on one atomic property per entry. Avoid combining multiple criteria into a single statement (e.g., separate "spelling" from "punctuation").
- Semantic Validity and Natural Fit: Do not force unnatural or meaningless combinations. Theoretical sources, softness degrees, and domains must align logically (e.g., do not pair a strictly rule-based Measurable mechanic with Degree 5 -- Very Soft, or Subjectivity with Degree 1 -- Measurable).
- Balanced Distribution: Maintain an even, balanced representation across all options in the three axes across the overall dataset.

Axis 1: Domain
Grammar, Usage, Mechanics, Style, Relevance, Organization, Development, Cohesion, Coherence, Thesis Clarity, Persuasiveness.

Axis 2: Source of Softness (Why Readers Disagree)
- Measurable (Rule-Based): Based on clear, exact rules or counts that have a simple right-or-wrong answer.
- Evaluative Vagueness (Cutoff Disagreement): Readers agree on what is being looked at, but disagree on where to draw the line for "enough."
- Underspecification (Unclear Scope): The general requirement is mentioned, but key details (such as exact placement or format) are left unstated.
- Multidimensionality (Multiple Factors): The trait is made up of several smaller parts, and different readers care more about different parts.
- Subjectivity (Personal Preference): Based purely on personal taste, feeling, or individual reader reaction; no single right answer exists.
- Essentially Contested (Fundamental Disagreement): Readers hold opposite core beliefs about what makes an argument valid or believable.

Axis 3: Degree of Softness (Reader Agreement Scale)
1 Not soft -- One interpretation. Anyone applying the description would arrive at the same reading.
2 Slightly soft -- One clear interpretation, with minor variation in details that would rarely change the outcome.
3 Moderately soft -- A few defensible interpretations, roughly two or three, with one of them somewhat more natural than the rest.
4 Very soft -- Several defensible interpretations, none of them clearly dominant.
5 Extremely soft -- Many defensible interpretations, spread widely, with no natural starting point.

Output Requirements: For each entry, provide:
- Domain
- Source of Softness
- Degree of Softness
- Atomic Attribute Description (a simple, everyday description of the trait)
- Everyday Explanation (a simple explanation of why readers agree or disagree)
- Examples of a couple of interpretations that could be made for the attribute description. This is a sanity check also whether the softness score is correct.
\end{lstlisting}
\end{promptbox}

\newpage
\subsubsection{Complete Description Pool}
Table~\ref{tab:appendix-descriptions} gives all descriptions and their construction metadata. Domain and source labels organize the stimulus pool.

\begin{table*}[t]
\centering
\footnotesize
\setlength{\tabcolsep}{4pt}
\renewcommand{\arraystretch}{0.92}
\caption{The 50 descriptions used in \prt{1} and \prt{2}, grouped by the dimension of essay
assessment they target. \emph{Deg.} is the softness degree assigned when the pool was
constructed}
\label{tab:appendix-descriptions}
\begin{minipage}[t]{0.49\textwidth}
\begin{tabular}[t]{@{}p{0.76\linewidth}ll@{}}
\toprule
\textbf{Description} & \textbf{Src.} & \textbf{Deg.} \\
\midrule
\multicolumn{3}{@{}l}{\textbf{Grammar}} \\
\cmidrule(r){1-3}
The essay uses correct verb forms. & \textsc{rule} & 1 \\
The essay uses complete sentences (no sentence fragments). & \textsc{rule} & 1 \\
The essay says ``it'' for one thing, and ``they'' when there's more than one. & \textsc{rule} & 1 \\
Sentences don't run on for too long. & \textsc{vague} & 2 \\
Pronoun references are unambiguous. & \textsc{scope} & 3 \\
\addlinespace[3pt]
\multicolumn{3}{@{}l}{\textbf{Usage}} \\
\cmidrule(r){1-3}
The essay avoids double negatives. & \textsc{rule} & 1 \\
Words are used correctly. & \textsc{vague} & 2 \\
Vocabulary matches the essay's subject area. & \textsc{scope} & 3 \\
The word choice paints a vivid picture for the reader. & \textsc{subj} & 4 \\
\addlinespace[3pt]
\multicolumn{3}{@{}l}{\textbf{Mechanics}} \\
\cmidrule(r){1-3}
Every sentence ends with proper punctuation. & \textsc{rule} & 1 \\
Proper nouns are capitalized. & \textsc{rule} & 1 \\
Paragraph spacing looks consistent throughout. & \textsc{vague} & 2 \\
Citations follow a consistent style. & \textsc{scope} & 3 \\
Punctuation choices, like dashes and colons, feel well-judged. & \textsc{subj} & 4 \\
\addlinespace[3pt]
\multicolumn{3}{@{}l}{\textbf{Style}} \\
\cmidrule(r){1-3}
Sentences vary in length. & \textsc{vague} & 2 \\
The essay has a formal style. & \textsc{multi} & 3 \\
The writing sounds confident. & \textsc{multi} & 4 \\
The writing is interesting to read. & \textsc{subj} & 4 \\
The writing shows creative flair. & \textsc{cont} & 5 \\
\addlinespace[3pt]
\multicolumn{3}{@{}l}{\textbf{Relevance}} \\
\cmidrule(r){1-3}
The essay's title matches its actual topic. & \textsc{rule} & 1 \\
Examples used in the essay relate to the topic. & \textsc{vague} & 2 \\
Every paragraph relates to the main topic. & \textsc{scope} & 3 \\
The examples used in the essay feel relatable to the reader. & \textsc{subj} & 4 \\
\bottomrule
\end{tabular}
\end{minipage}%
\hfill
\begin{minipage}[t]{0.49\textwidth}
\begin{tabular}[t]{@{}p{0.76\linewidth}ll@{}}
\toprule
\textbf{Description} & \textbf{Src.} & \textbf{Deg.} \\
\midrule
\multicolumn{3}{@{}l}{\textbf{Organization}} \\
\cmidrule(r){1-3}
The essay has an introduction, body, and conclusion. & \textsc{rule} & 1 \\
Paragraphs are arranged in a logical order. & \textsc{vague} & 2 \\
Each paragraph opens with a topic sentence. & \textsc{scope} & 3 \\
Paragraphs build on each other. & \textsc{multi} & 3 \\
The essay's structure feels natural rather than rigid. & \textsc{cont} & 5 \\
\addlinespace[3pt]
\multicolumn{3}{@{}l}{\textbf{Development}} \\
\cmidrule(r){1-3}
Each claim has some evidence behind it. & \textsc{vague} & 2 \\
Ideas are explored, not just listed. & \textsc{scope} & 3 \\
Ideas are explained in enough detail. & \textsc{multi} & 4 \\
The essay offers an original insight, not just familiar facts. & \textsc{subj} & 4 \\
The essay shows deep analysis rather than surface-level summary. & \textsc{cont} & 5 \\
\addlinespace[3pt]
\multicolumn{3}{@{}l}{\textbf{Cohesion}} \\
\cmidrule(r){1-3}
The essay uses pronouns that clearly point back to a specific noun. & \textsc{rule} & 1 \\
It's always clear what ``it'' or ``they'' is referring to. & \textsc{vague} & 2 \\
Sentences connect smoothly to each other. & \textsc{scope} & 3 \\
Related ideas are linked across paragraphs. & \textsc{multi} & 4 \\
\addlinespace[3pt]
\multicolumn{3}{@{}l}{\textbf{Coherence}} \\
\cmidrule(r){1-3}
The connection between sentences is easy to follow. & \textsc{scope} & 3 \\
The essay's parts fit together well. & \textsc{multi} & 4 \\
Transitions between sections feel natural. & \textsc{subj} & 4 \\
The essay makes sense as a whole. & \textsc{cont} & 5 \\
\addlinespace[3pt]
\multicolumn{3}{@{}l}{\textbf{Thesis Clarity}} \\
\cmidrule(r){1-3}
The essay states its main point in the first paragraph. & \textsc{rule} & 1 \\
The essay's main argument is stated in a single sentence. & \textsc{rule} & 1 \\
The main point of the essay is clear. & \textsc{vague} & 2 \\
The main point fits the length of the essay. & \textsc{scope} & 3 \\
The main point clearly states the writer's side and their key reasons for it. & \textsc{multi} & 4 \\
\addlinespace[3pt]
\multicolumn{3}{@{}l}{\textbf{Persuasiveness}} \\
\cmidrule(r){1-3}
The reasons given lead logically to the conclusion. & \textsc{scope} & 3 \\
The essay builds a strong case. & \textsc{multi} & 4 \\
The argument is convincing. & \textsc{cont} & 5 \\
The essay's claims are ethically responsible. & \textsc{cont} & 5 \\
\bottomrule
\end{tabular}
\end{minipage}
\end{table*}



\subsection{Coding Details}
\label{app:softness-coding}
Two coders independently grouped the responses to each description by their operational meaning. Wording differences alone did not warrant a separate group. 
Responses belonged together when they implied equivalent criteria for deciding which essays satisfied the description.
The coding guidance instructed coders to start a group with the first response and consider subsequent responses against the groups already formed.
A response warranted a new group if it introduced a different idea or committed to a particular procedure that an existing group left open.
Unique responses, and responses too vague to assign to an existing group, received separate groups. 
Responses completely unrelated to the description were marked as outliers. 
Coders could record notes about difficult decisions.

The coders flagged 19 and 16 responses as unrelated, with seven overlapping flags. 
The analysis removed the union of these flags from both partitions, excluding 28 responses and retaining 592. 
The independent partitions were not consolidated. 
Mean Adjusted Rand Index across descriptions was .242.
Absolute-agreement reliability of the two-coder mean, ICC(A,2), was .814 for raw group counts and .792 for rarefied counts. 
Agreement in the number of groups therefore does not imply agreement about the identity of every reading.

\section{Agent Framework}
\label{app:agent-framework}

\subsection{Softness Calibration}
\label{app:calibration}
The workflow uses two cutoffs to determine whether to generate functions and
whether to continue clarifying the current topic. Table~\ref{tab:appendix-gates}
states the policy described in the main paper, including its boundary cases.
\begin{table}[htbp]
\centering\small
\caption{Softness measure calibration for \app}
\label{tab:appendix-gates}
\begin{tabular}{@{}lp{.28\linewidth}p{.42\linewidth}@{}}
\toprule
Score $s$ & Function generation & Next interaction \\
\midrule
$s\geq65$ & No & Clarification \\
$30<s<65$ & Yes & Clarification after proposal review \\
$s\leq30$ & Yes & Invitation for a suggestion or a new topic \\
\bottomrule
\end{tabular}
\end{table}

The intended distinction is between descriptions requiring further
disambiguation before a proposal can be made, descriptions concrete enough
for provisional proposals but still open to clarification, and descriptions
for which the user can elect to move on. These are interaction-policy
decisions, not measures of the correctness of a generated function.


\subsection{Technical Report}
\label{app:technical-report}
\label{app}

\subsubsection{Softness Computation}
The current implementation of the our measures uses \texttt{gemini-3.5-flash} for restatement and \texttt{gemini-3.5-flash-lite} for interpretation generation. Restatement uses temperature 0; interpretation generation uses temperature .9 and
requests ten candidates. 
The restatement retains the user's constraints, gives precedence to later revisions, removes retracted requirements and adds no new measurement criteria. 
Interpretation generation starts afresh from the current specification after each turn.
Each candidate contains an attribute label and, in the full-output mode, a concrete description.
The score embeds the \emph{labels}, using \texttt{sentence-transformers/all-MiniLM-L6-v2}. 
It does not embed generated Python function code. 

Let $K$ denote their cosine-similarity matrix. The available implementation symmetrizes $K$, clips its entries to $[0,1]$, fixes its diagonal to one, and computes eigenvalues of $K/n$. 
It retains eigenvalues greater than $10^{-12}$ and computes

\begin{equation}
 V=\exp\!\left(-\sum_{\lambda_i>10^{-12}}
                 \lambda_i\log\lambda_i\right).
\end{equation}

\noindent In the default raw-Vendi mode, the score is the rounded, clipped linear map

\begin{equation}
 s=\operatorname{round}\!\left(
 \min\!\left(100,\max\!\left(0,100\frac{V-1}{10-1}\right)\right)\right).
\end{equation}

\noindent The bounds specify score scaling; the workflow cutoffs specify actions.


\subsubsection{Function Generation and Execution}
The Function Agent receives the current discussion block's question--answer pairs and translates them into numerical Python attributes using the shared CodeAct service. As described in the paper, candidates undergo AST validation and sample execution, with failures returned to the agent for revision until success or exhaustion of the retry budget. 
Users review descriptions of the resulting computations and can accept, reject or refine the proposals. 
Successful parsing or sample execution does not establish that the function faithfully represents the intended criterion.


\subsubsection{Dataset and Projection}
Accepted attributes populate an essay-by-attribute table. 
The implementation described in the paper uses scikit-learn PCA and DBSCAN to display items according to their current attribute values. 
Open-ended question generation can receive essay excerpts sampled from the largest non-noise cluster, along with the existing attributes and interaction history.


\subsubsection{Prompt Specifications}
The following templates retain the recoverable prompt wording, with variable context represented by named placeholders. 
Static domain substitutions use ``essay'' and ``essays.'' 
These are source templates rather than examples of participant responses. 

\smallparagraph{Restating the Attribute Specification} After each user message, the system consolidates the conversation so far into a single specification.
\begin{promptbox}[Restatement Prompt]
\textbf{\footnotesize System prompt}
\vspace{2pt}
\begin{lstlisting}[style=promptcode]
An ATTRIBUTE is a property of essays in a corpus that a user wants to measure or score across that corpus. The user describes the attribute in natural language; your job is to help turn that description into a concrete decision procedure that can be applied to every essay. Descriptions vary from precise (one clear way to measure) to ambiguous (many plausible operationalizations).

The user has described the attribute over several messages. Your job is to write down the attribute as it currently stands: one consolidated specification that a reader who never saw the conversation could use.

Rules:
1. Integrate every message. When a later message conflicts with an earlier one, the later message wins. When the user retracts something ("forget that", "scratch that", "that's too strict"), drop the retracted constraint entirely.
2. Keep every constraint the user actually stated: thresholds, counts, positions, required words, scope (sentence / paragraph / whole essay).
3. Keep the user's hedges and explicit ceilings. Phrases like "any visible link is enough", "however that happens", "nothing more than that", "that's the whole bar" limit how far the attribute goes; preserve their effect.
4. Add nothing. No thresholds, units, examples, procedures, or sub-aspects the user did not mention. If the user's wording is vague, the specification stays vague. Reuse the user's own key terms rather than substituting more technical ones.
5. Remove conversational framing: "I mean", "actually", "honestly", "yep", acknowledgements, and any reference to earlier messages or to the conversation itself.
6. Write one to two sentences describing the property of essays being measured. Do not make it longer than the user's messages combined. Do not list alternatives.

Output strict JSON: {"specification": "..."}. No preamble.
\end{lstlisting}
\vspace{4pt}
\textbf{\footnotesize User prompt}
\vspace{2pt}
\begin{lstlisting}[style=promptcode]
The user's messages, in order:

Message 1 (initial description):
{initial_description}

Message 2 (added later):
{later_user_message}

Message 3 (latest):
{latest_user_message}

Write the consolidated specification as JSON.
\end{lstlisting}
\end{promptbox}


\smallparagraph{Enumerating Interpretive Readings} Given the current specification, the system enumerates distinct operational readings of the description.
\begin{promptbox}[Interpretation Prompt]
\textbf{\footnotesize System prompt}
\vspace{2pt}
\begin{lstlisting}[style=promptcode]
An ATTRIBUTE is a property of essays in a corpus that a user wants to measure or score across that corpus. The user describes the attribute in natural language; your job is to help turn that description into a concrete decision procedure that can be applied to every essay. Descriptions vary from precise (one clear way to measure) to ambiguous (many plausible operationalizations).

Your job in this step is to enumerate the operational READINGS of the user's attribute description -- the ways a faithful, careful reader could turn that description into a concrete decision procedure over essays.

For each reading, produce a structured object with exactly two fields:
- attribute_label: a 3-8 word NOUN PHRASE naming the property this reading measures. This is a NAME, not a sentence -- no subject, no verb, no procedural framing, no 'is/has/uses/contains/checks'. Think of it as a column header or variable name. CRITICAL: do NOT include the word 'essay' or 'essays' in the label -- the domain is already implied.
    BAD  (procedural, templated):
      - 'essay uses emotional adjectives'
      - 'essay contains autobiographical content'
      - 'word count of the essay exceeds 500'
      - 'checks for narrative tension'
    GOOD (nominal, content-bearing):
      - 'emotional adjective density'
      - 'autobiographical childhood scenes'
      - 'word count above 500'
      - 'presence of narrative tension'
  Labels for closely-related readings can share head nouns ('word count above 500', 'word count above 1000'); that is fine. What must NOT happen is every label sharing the same sentence template.
- full_description: a self-contained concrete attribute description -- ONE decision procedure, no bundled alternatives, no 'or'/slashes -- that could be handed off to a function generator (e.g. 'essay uses at least three distinct emotional adjectives' rather than 'is emotional'). Procedural framing belongs HERE, not in attribute_label.

Stay faithful to what the user actually said. Each reading must be one a reasonable analyst would actually consider -- not a contrived variation invented to fill space. The number of items requested is the BUDGET for enumeration, not a target for forced diversity.

Each reading must satisfy every clarification the user has given. Reject readings that contradict any of the user's answers.

The natural consequence of fidelity:
- When the description leaves room for genuinely divergent readings, the items will span different angles: different units of analysis, different evidence types, different thresholds, different sub-aspects of the concept, different scopes (sentence / paragraph / whole essay), different stakeholder perspectives.
- When the description tightly pins down the measurement, the items will converge on the same core idea; the variation between them will be small and incidental -- slight differences in threshold value, in framing, in scope of evidence. That is the correct outcome. Do not manufacture distinctions the description does not license.

OUTPUT: strict JSON -- a single array of objects, each with exactly `attribute_label` and `full_description`. No preamble, no commentary, no markdown code fences. Just the JSON array.
\end{lstlisting}
\vspace{4pt}
\textbf{\footnotesize User prompt}
\vspace{2pt}
\begin{lstlisting}[style=promptcode]
The user's description of the attribute, in their own words:
"""
{current_specification_or_description}
"""

Produce exactly 10 readings of the user's attribute description as a JSON array. Each object MUST have:
- attribute_label: a 3-8 word NOUN PHRASE naming the property -- NOT a sentence. No subject, no verb, no procedural framing, and do NOT include 'essay' or 'essays' in the label (e.g. 'emotional adjective density', 'word count above 500' -- NOT 'essay uses emotional adjectives').
- full_description: a self-contained concrete attribute over essays (one decision procedure, no bundled alternatives). Procedural framing belongs here, not in attribute_label.

Stay faithful to what the user said. The 10 readings should be the ones a thoughtful analyst would actually consider; do not pad with contrived variations.

- If the description licenses many different angles, the 10 items will naturally span them -- different units, evidence types, thresholds, sub-aspects, scopes (sentence / paragraph / whole essay), stakeholder perspectives.
- If the description tightly fixes the measurement, the 10 items will cluster around the same core idea; their labels will be near each other and their descriptions will differ only in incidental ways (a threshold value, a framing choice, the scope of evidence). Either outcome is correct -- the count is a budget, not a diversity target.

Output ONLY the JSON array -- exactly 10 objects, no more, no fewer, no surrounding prose, no markdown fences.
\end{lstlisting}
\end{promptbox}

\smallparagraph{Open-Ended Question Generation} To probe breadth of preference across many dimensions rather than depth on one, the system generates a single open-ended question at each turn.
\begin{promptbox}[Open-Ended Question Prompt]
\begin{lstlisting}[style=promptcode]
# Role
You are an expert interviewer for preference elicitation.

# Task
Your task is to learn what characteristics and attributes a user prefers in student essays. Essay preferences can vary widely, so you should seek to understand their preferences across many dimensions; in other words, go for breadth rather than depth. Do not assume a user has given a complete answer to any question, so make sure to keep probing for what attributes they prefer.
The implemented system is: user's essay attributes

# Context
Previous questions and answers:
{interaction_history}
{attribute_context}


Cluster-separation evidence (sampled essays):
- Source: random sample from the largest non-noise projection cluster in the current session.
- Purpose: use these examples to ask a question that helps distinguish this cluster from other essays.
- Format: each line is "Essay <id> (cluster <id>): <excerpt>".
{cluster_essay_context}

Additional instructions:
{additional_query_note_if_configured}


# Objective
Generate one open-ended question that is maximally informative about the user's desired behavior, beyond what has already been asked.

# Constraints
- Ask about a different aspect than earlier questions.
- Keep the question bite-sized and focused (do not ask for too much at once).
- If cluster essay context is provided, use it to improve cluster separability.
- Do not overfit to the specific sampled essays.

# Output format
- Return exactly one open-ended question.
- Return plain text only.
- Do not include explanation, bullets, or any extra content.

Now generate the question:
\end{lstlisting}
\end{promptbox}


\smallparagraph{Clarification Question Generation} Following the user's most recent answer, the system generates a follow-up clarifying question targeted at whatever remains vague or underspecified.
\begin{promptbox}[Clarification Question Prompt]
\begin{lstlisting}[style=promptcode]
Your task is to learn what characteristics and attributes a user prefers in student essays. Essay preferences can vary widely, so you should seek to understand their preferences across many dimensions. Do not assume a user has given a complete answer to any question, so make sure to keep probing for what attributes they prefer.

Previous questions:
{interaction_history}

Generated function titles so far:
{generated_function_titles}

Generate one concise clarifying question focused on extracting deeper insight from the user's most recent answer. The goal is to better understand exactly what they meant and what they prioritized, for the user's essay preferences. Use the generated function titles to identify what is vague, underspecified, or weakly related, and ask about one specific missing detail that would improve those functions. Avoid repeating earlier questions, and keep it answerable in 1-3 sentences. {additional_query_note_if_configured}
Generate only the clarifying question:
\end{lstlisting}
\end{promptbox}

\section{User Study}
\label{app:user-study}

\subsection{Participant Details}
\label{app:user-study-details}

We recruited 41 participants through Prolific, of whom 32 completed both conditions and passed the attention check, and are included in the analysis. 
The 32 participants (17 male, 15 female) were predominantly aged 25--34 (53\%, $n = 17$), with seven aged 18--24, three aged 35--44, two aged 45--54, one aged 55--64, and two aged 65 or older.
Nearly all were frequent LLM users: 59\% reported daily use and 38\% regular use, with only one occasional user.
The most commonly reported LLM tasks were obtaining feedback on writing (69\%), essay writing (66\%), grading (53\%), and other writing tasks (53\%).
Participants reported moderate-to-high experience with both writing (argumentative: $M = 3.55$, $SD = 0.99$; narrative: $M = 3.71$, $SD = 0.97$; both on a 5-point scale) and evaluation (argumentative: $M = 3.52$, $SD = 1.18$; narrative: $M = 3.61$, $SD = 1.02$).




\subsection{Procedure}
\label{app:user-study-procedure}
After consent and background questions, participants completed a general tutorial about translating preferences into attributes, using movies as a separate example domain. 
A comprehension check asked the main purpose of the tool, with the intended answer ``To help you identify and articulate what you value in a good essay.'' 
The local protocol routes other answers to an attention-failure exit. 
Tool-specific checks explain that accepting a card indicates relevance to one's preferences, rather than verification that the measurement is correct.

Before each block, participants received the assigned dataset introduction
and tool tutorial. T
he opening question was fixed by essay genre and shared
across conditions.

\begin{quote}
\textbf{Argumentative essays.} What makes an argumentative essay good in
your view? Describe something that matters to you in your own words.

\textbf{Narrative essays.} What makes a narrative essay good in your view?
Describe something that matters to you in your own words.
\end{quote}

Participants were instructed to work for the full 15-minute block while a
timer displayed the remaining time. 
They could articulate preferences and review proposed attributes. 
Before the second block, the protocol asked them to develop different attributes for the new collection instead of reusing their first list.

At the end of each block, participants reviewed their attributes and answered the condition questionnaire, a task-completion question and a written reflection. 
The protocol also includes six workload ratings and, when a rejected proposal was available, a replay asking why it had been rejected. 
They then stated their preferred tool, explained their choice and compared what they could express with each tool.


\subsection{Questionnaires and Response Anchors}
\label{app:user-study-questionnaire}
The following wording was recovered from the local study protocol, version \texttt{2026-09-09}. 
Participants were instructed to answer according to what they wanted the attributes to capture by the end of each block, allowing their intended specifications to develop during the interaction.
Agreement responses were coded from 1 to 5 as ``Strongly disagree'', ``Somewhat disagree'', ``Neither agree nor disagree'', ``Somewhat agree'', ``Strongly agree.'' The mismatch item additionally offered ``Not applicable. 
I did not encounter a mismatch.'' as a nonnumeric response.

\begin{enumerate}
\item I could express the distinctions that mattered to me.
\item The final attribute definitions reflect what I wanted them to capture.
\item I understand how the tool translated my descriptions into measurable criteria.
\item The interaction helped me clarify what I wanted the attributes to capture.
\item The interaction helped me identify relevant aspects I had not initially considered.
\item I could guide the interaction in the direction I wanted.
\item I felt that the final attribute definitions reflected my own choices.
\item When a proposal did not match what I wanted, I could get the tool to address the mismatch.
\item Developing the attributes with this tool required substantial mental effort.
\item I spent effort on interaction steps that did not help me develop the attributes.
\end{enumerate}
Higher scores on the last two items indicate greater effort. The items describe perceived support and are reported individually; they are not direct tests of implementation fidelity.

\subsubsection{Workload}
The six workload sliders range from 0 to 100 in steps of five. All use Low and High anchors except Performance, which uses Good at 0 and Poor at 100.
\begin{description}
\item[Mental Demand.] How much mental and perceptual activity was required (e.g. thinking, deciding, calculating, remembering, looking, searching, etc.)? Was the task easy or demanding, simple or complex, exacting or forgiving?
\item[Physical Demand.] How much physical activity was required (e.g. pushing, pulling, turning, controlling, activating, etc.)? Was the task easy or demanding, slow or brisk, slack or strenuous, restful or laborious?
\item[Temporal Demand.] How much time pressure did you feel due to the rate or pace at which the tasks or task elements occurred? Was the pace slow and leisurely or rapid and frantic?
\item[Performance.] How successful do you think you were in accomplishing the goals of the task set by the experimenter (or yourself)? How satisfied were you with your performance in accomplishing these goals?
\item[Effort.] How hard did you have to work (mentally and physically) to accomplish your level of performance?
\item[Frustration Level.] How insecure, discouraged, irritated, stressed and annoyed versus secure, gratified, content, relaxed and complacent did you feel during the task?
\end{description}
No pairwise workload-weighting task is specified in the recovered protocol.

\subsubsection{Completion, Reflection and Comparison}
\begin{description}
\item[Completion.] At the end of the allotted time, did you feel you had finished developing the attributes you wanted? Responses were ``Yes'', ``Partly'', ``No''. An optional text field asked ``Why?
\item[Reflection.] Describe one moment when the tool helped or hindered you in developing the attributes. What happened, and how did it affect what you wanted? If no particular moment stands out, please say so.
\item[Tool preference.] Which tool would you choose for another attribute-construction task like this one? Responses were ``Tool A'', ``Tool B'', ``No preference''.
\item[Reason.] What is the main reason for your choice? For no preference, the prompt was ``If you selected No preference, please explain why.''
\item[Differences.] Was there anything you could express or explore with one tool that was difficult with the other? Please give an example, or say that you noticed no difference.
\end{description}


\subsection{Baseline}
\label{app:user-study-baseline}
Both conditions provide conversational input and access to the same function-generation service, with explicit review of proposed attributes.
The baseline sends the running conversation to its conversational assistant and produces attribute proposals after each user turn. 
It does not use
softness-gated generation or the sequence of \textit{open-ended}, \textit{clarification} and \textit{suggestion} phases. 
\app generates from the active discussion block and uses its score to select the next action. 
The comparison therefore concerns the complete workflows, including their context-management policies.

The local tutorial shows the attribute table and projection in both
conditions, with an additional interaction-history panel in \app.

\subsubsection{Baseline Conversation Prompt}

For the baseline condition, the system replies as an unstructured conversation partner rather than following the restate/interpret/clarify pipeline of \app.

\begin{promptbox}[Baseline Prompt]
\textbf{\footnotesize System prompt}
\vspace{2pt}
\begin{lstlisting}[style=promptcode]
You are a helpful assistant talking with a user about user's essay preferences.

The user is working through what they care about in student essays. Your job is to be a useful conversation partner while they do that: respond to what they actually said, help them think it through, and let them steer.

Guidelines:
- Reply in plain conversational prose. No headers, no bullet lists, no numbered plans.
- Keep replies short, at most three or four sentences.
- Follow the user's lead. If they ask you something, answer it. If they think out loud, engage with the substance.
- You may ask a follow-up when it is genuinely useful, but you are not running an interview and you do not have to end every turn with a question.
- Do not summarize the conversation back to the user unless they ask.
- Do not talk about attribute functions, scoring code, or the cards shown elsewhere in the interface, unless the user brings them up first.
\end{lstlisting}
\vspace{4pt}
\textbf{\footnotesize User prompt}
\vspace{2pt}
\begin{lstlisting}[style=promptcode]
Conversation so far:
{complete_running_transcript}

Write the assistant's next reply. Return the reply text only, with no speaker label and no surrounding quotes.
\end{lstlisting}
\end{promptbox}

\subsection{Analysis Definitions and Supplementary Results}
\label{app:user-study-analysis}
Process measures distinguish functions generated from output attributes.
A function generation produces a single function candidate. 
Refinement changes an existing attribute; a revised attribute has more than one stored version. 
Acceptance counts refer to stored final properties that are collected in the table. 
Service-call duration measures a recorded generation request, rather than the participant's complete waiting
time. 


\subsubsection{Paired Results}
The saved 32-participant process inventory provides the descriptive counts
in Table~\ref{tab:results}. Counts summarize recorded output and decisions;
they do not assess measurement quality or intention--implementation alignment.

\begin{table}[htbp]
\centering\small
\caption{Descriptive process counts across 32 completed participants per
condition. Accepted slots use stored final status, subject to the decision
discrepancy noted above. Refinements exclude regeneration.}
\label{tab:results}
\begin{tabular}{@{}lrr@{}}
\toprule
Measure & \app & Baseline \\
\midrule
Attribute slots & 283 & 357 \\
Stored accepted slots & 252 & 324 \\
Explicit refinement events & 4 & 17 \\
Participants making an explicit refinement & 2 & 10 \\
\bottomrule
\end{tabular}
\end{table}



\end{document}